\documentclass[11pt,document,nofootinbib,superscriptaddress,onecolumn,preprintnumbers,balancelastpage]{article}
\pdfoutput=1
\usepackage{jheppub}
\usepackage[T1]{fontenc} 

\usepackage{amssymb}
\usepackage{amsmath}
\usepackage{epsfig}
\usepackage{comment}
\usepackage{color}
\usepackage{multirow}
\usepackage{capt-of}
\usepackage{siunitx}
\usepackage{graphicx}
\usepackage{gensymb}
\usepackage{slashed}
\usepackage{tabularx}
\usepackage{enumitem}
\usepackage{multirow}
\usepackage{booktabs}
\usepackage{isotope} 
\usepackage{xcolor}
\usepackage{bm}
\usepackage{array,mathtools}

\newcolumntype{C}{>{$}c<{$}}
\AtBeginDocument{
\heavyrulewidth=.08em
\lightrulewidth=.05em
\cmidrulewidth=.03em
\belowrulesep=.65ex
\belowbottomsep=0pt
\aboverulesep=.4ex
\abovetopsep=0pt
\cmidrulesep=\doublerulesep
\cmidrulekern=.5em
\defaultaddspace=.5em
}

\usepackage{tikz}
\usepackage{tkz-euclide}
\usetkzobj{all}
\usetikzlibrary{backgrounds}
\usetikzlibrary{decorations.pathmorphing}	
\tikzset{
    v/.style={decorate, decoration={snake, segment length=3mm, amplitude=0.75mm}, draw},
    f/.style={draw=black, postaction={decorate},
        decoration={markings,mark=at position .6 with {\arrow[very thick]{latex}}}},
    fb/.style={draw=black, postaction={decorate},
        decoration={markings,mark=at position .4 with {\arrowreversed[very thick]{latex}}}},
    fnar/.style={draw=black},
    g/.style={decorate, draw=black,
        decoration={coil,amplitude=3pt, segment length=3.5pt}},
    s/.style={dashed,draw=black, postaction={decorate},
        decoration={markings,mark=at position .55 with {\arrow[very thick]{latex}}}},
    sb/.style={dashed,draw=black, postaction={decorate},
        decoration={markings,mark=at position .55 with {\arrowreversed[draw=black,very thick]{latex}}}},
    snar/.style={dashed,draw=black,line width =1.25pt},
}

\newcommand{\MM}{\mathcal{M}}

\newcommand{\h}[1]{\isotope[#1]{H}}
\newcommand{\he}[1]{\isotope[#1]{He}}

\newcommand{\exc}{{(*)}}

\definecolor{myred}{rgb}{1., 0., 0.0014072231026931448}
\definecolor{myorange}{rgb}{1., 0.6411371983743057, 0.}
\definecolor{myyellow}{rgb}{0.9998538461538462, 0.8472174696143701, 0.}
\definecolor{mylime}{rgb}{0.8296461722852855, 0.9416163076923076, 0.}
\definecolor{mygreen}{rgb}{0.14471625664756718, 0.9071012307692308, 0.2631632274752619}
\definecolor{myaqua}{rgb}{0.05406425942348916, 0.8614636923076924, 0.7734102885268764}
\definecolor{mysky}{rgb}{0., 0.5883488108126352, 0.9445016153846155}
\definecolor{myblue}{rgb}{0., 0.34480560463730825, 1.}
\definecolor{mynavy}{rgb}{0.25166566047369227, 0., 0.8886266153846154}
\definecolor{mypurple}{rgb}{0.6303632859199042, 0., 0.7270160000000001}
\definecolor{mydrkpurp}{rgb}{0.4335277943813507, 0., 0.5}

\newcommand{\prn}[1]{ \left(  #1 \right) }

\newcommand{\avg}[1]{\left< #1 \right>}

\newcommand{\ord}[1]{\mathcal{O}\left(#1 \right)}

\newcommand{\magn}[1]{\left| #1 \right|}

\newcommand{\al}[1]{\begin{align} #1 \end{align}}

\newcommand{\mNj}{M_{A_j,Z_j}}

\newcommand{\mCCth}{m_{\text{th}}^{\beta}}
\newcommand{\mCCthj}{m_{\text{th},\, j}^{\beta}}

\newcommand{\mNprimePlus}{M_{A, Z-1}}

\title{Absorption of Fermionic Dark Matter \\ by Nuclear Targets}

\author[1,2]{Jeff A. Dror, }
\author[3]{Gilly Elor, }
\author[2,1]{and Robert McGehee}
\affiliation[1]{Theory Group, Lawrence Berkeley National Laboratory, Berkeley, CA 94720, USA}
\affiliation[2]{Berkeley Center for Theoretical Physics, University of California, Berkeley, CA 94720, USA}
\affiliation[3]{Department of Physics, University of Washington, Seattle, WA 98195, U.S.A.}
 
\abstract{Absorption of fermionic dark matter leads to a range of distinct and novel signatures at dark matter direct detection and neutrino experiments. We study the possible signals from fermionic absorption by nuclear targets, which we divide into two classes of four Fermi operators: neutral and charged current. In the neutral current signal, dark matter is absorbed by a target nucleus and a neutrino is emitted. This results in a characteristically different nuclear recoil energy spectrum from that of elastic scattering. The charged current channel leads to induced $\beta$ decays in isotopes which are stable in vacuum as well as shifts of the kinematic endpoint of $ \beta$ spectra in unstable isotopes. To confirm the possibility of observing these signals in light of other constraints, we introduce UV completions of example higher dimensional operators that lead to fermionic absorption signals and study their phenomenology. Most prominently, dark matter which exhibits fermionic absorption signals is necessarily unstable leading to stringent bounds from indirect detection searches. Nevertheless, we find a large viable parameter space in which dark matter is sufficiently long lived and detectable in current and future experiments.}

\begin{document} 
\maketitle
\flushbottom

\section{Introduction}
Despite increasing sensitivities, dark matter direct detection experiments have continued to yield null results for the long sought after weakly interacting massive particle (WIMP) \cite{Akerib:2016vxi,Tan:2016zwf,Aprile:2017iyp}. As such, both theoretical and experimental programs have moved beyond the WIMP paradigm. On the theoretical side, explorations of alternative dark matter candidates and production mechanisms have motivated dark matter masses below a GeV~\cite{Griest:1990kh,Pospelov:2007mp,Hochberg:2014dra,Hochberg:2014kqa,Kuflik:2015isi,Carlson:1992fn,Pappadopulo:2016pkp,Farina:2016llk,Dror:2016rxc,Dror:2017gjq,Hall:2009bx,Cheung:2010gj,Cheung:2010gk}. 
On the experimental side, progress has been made on the size frontier --- where ton-scale experiments are needed to search for signals with rates consistent with current bounds. 
In tandem, new proposals hope to probe regions of parameter space interesting for lighter dark matter candidates~\cite{Essig:2011nj,Graham:2012su,Essig:2012yx,Essig:2015cda,Hochberg:2016ntt,Derenzo:2016fse,Essig:2017kqs,Budnik:2017sbu,Cavoto:2017otc,Kurinsky:2019pgb,Hochberg:2015pha,Hochberg:2015fth,Hochberg:2016ajh,Schutz:2016tid,Knapen:2016cue,Hochberg:2017wce,Knapen:2017ekk,Szydagis:2018wjp,Baryakhtar:2018doz,Griffin:2018bjn}. 

Another strategy for progress is to explore novel dark matter direct detection signals (see e.g.,~\cite{Kile:2009nn,Agashe:2014yua,Lasserre:2016eot}). With this direction in mind, we recently considered the absorption of a fermionic dark matter particle in a detector~\cite{Dror:2019onn} --- a scenario in which the dark matter mass energy is available to the target (in contrast to the well studied elastic scenario where only its velocity-suppressed kinetic energy may be imparted).  In our present work, we build on this foundation by exploring and cataloging all such absorption signals off nuclear targets, leaving the consideration of electron targets to future work~\cite{NCelectrons}.~\footnote{See~\cite{Fiaschi:2019evv} for recent work on MeV sterile neutrino dark matter elastically scattering off electrons.} We organize these signals by the corresponding types of higher dimensional operators which lead to fermionic absorption.

We begin by considering signals from absorption of fermionic dark matter from ``neutral current'' processes of the form 
\begin{align} 
\overset{\textbf{\fontsize{2pt}{1pt}\selectfont(---)}}{\chi }  + \isotope[A][Z]{X}\rightarrow  \overset{\textbf{\fontsize{2pt}{1pt}\selectfont(---)}}{\nu  }  + \isotope[A][Z]{X} \, ,
\end{align} 
where \isotope[A][Z]{X} is the nuclear target with atomic number $Z$ and atomic mass number $A$, $\chi$ is the dark matter, and $\nu$ is a Standard Model (SM) neutrino. This signal is generated by dimension-6 neutral current operators of the form $\left[\bar{\chi} \Gamma _i  \nu \right] \left[ \bar{n} \Gamma _j n \right]$ and $\left[\bar{\chi} \Gamma _i  \nu \right] \left[ \bar{p} \Gamma _j p \right]$, where $ \Gamma _i = \left\{ {\mathbf{1}} , \gamma _5 , \gamma _\mu , \gamma _\mu \gamma _5 ,   \sigma _{\mu\nu} \right\}$ contains all possible Lorentz structures. Since dark matter must be lighter than the nucleons to avoid rapid decays, energy momentum conservation ensures that the outgoing neutrino carries away most of the dark matter (mass) energy. Nevertheless, a fraction of the $ \chi $ mass is still converted into kinetic energy for the recoiling nucleus, resulting in a distinct signal. Like spin-independent WIMP scattering, the absorption rates can enjoy a coherent enhancement for larger nuclei. In this work, we study the neutral current in detail by surveying current experiments and discussing the types of future experiments best suited to detect these processes. 

Another set of fermionic absorption signals are induced $\beta$ decays
\begin{align} 
\overset{\textbf{\fontsize{2pt}{1pt}\selectfont(---)}}{\chi }+\isotope[A][Z]{X}  \rightarrow e ^{ \pm }+\isotope[A][Z\mp1]{X}^\exc  \,,
\end{align} 
where $(*)$ denotes a possible excited state of the nucleus (which range from below an MeV to $ 10 $s of MeV above the ground state depending on the isotope). Such ``charged current'' processes are generated by dimension-6 operators of the form $\left[\bar{ \chi} \Gamma _i e \right] \left[\bar{n } \Gamma _j  p \right]$. The induced decay can occur in isotopes that are stable or unstable in the vacuum. Stable (or meta-stable) isotopes exist in macroscopic quantities in current experiments and so these can be employed to look for multiple correlated signals: the energetic ejected $e^\pm$, the recoil of the daughter nucleus, a $\gamma$ from the decay of the excited daughter nucleus, and another $\beta$ decay of the final nucleus, if it is unstable. Due to these multiple signals and large $ e^\pm $ (and potentially photon) energy, dedicated searches for induced $\beta$ decays do not rely on the nuclear recoils being above a given experimental threshold. However, these processes themselves have kinematic thresholds allowing them to only probe dark matter masses larger than $\sim 400 \text{ keV}$. This kind of signal has been considered in the context of sterile neutrino dark matter detection~\cite{Lasserre:2016eot}, where it was concluded that immense quantities of Dysprosium (which is rare but has an anomalously small $ \beta $ decay energy threshold of $ \sim 2.5~{\rm keV} $) is needed to probe the parameter space consistent with indirect detection bounds from sterile neutrino decay. In this paper, we study alternative dark matter candidates, and find that current experiments can easily observe signals consistent with other constraints.

In principle, one can look for induced $ \beta ^- $ or $ \beta ^+ $ decays for all isotopes within a detector. Indeed, we find many induced $\beta^-$ decay targets in current dark matter direct detection and neutrino experiments. However, induced $  \beta ^+ $ decay rates suffer relative to those of $\beta^-$ due to the Coulomb repulsion of the $e^+$ by the nucleus as well as Pauli blocking effects of the outgoing neutron. As such, we will primarily be interested in signals from induced $\beta^+$ decays off of Hydrogen targets in neutrino experiments, as considered in~\cite{Kile:2009nn} for Super-Kamiokande. Nonetheless, induced $\beta^+$ decays are worth consideration since they allow complementary isotope targets in experiments to probe the same operators and might be necessary to search for the asymmetric dark matter scenario, where only $ \chi $ or $ \overline{ \chi }  $ may be present today. In this work, we survey the current experiments which can be used to look for induced $ \beta $ decays of stable isotopes and their projected reach. 

For unstable isotopes, it is more challenging to accumulate macroscopic targets in detectors. Nevertheless, since they have no induced $\beta$ decay thresholds, they may be used to detect arbitrarily light dark matter. To find these signals, one can look for outgoing $\beta$ with energies beyond the kinematic endpoint of the target isotope's $\beta$ decay spectra. Despite these practical challenges, there are proposals with other primary physics goals which rely on $\beta^-$ decaying isotopes, such as PTOLEMY~\cite{Betts:2013uya,Baracchini:2018wwj}. Previous studies have focused on sterile neutrinos where it's challenging to compete with current decay bounds~\cite{Li:2010vy,Long:2014zva}. We propose to test light dark matter using this signal and overview the types of experiments necessary to probe parameter space consistent with other constraints. 

Importantly, dark matter candidates which allow either the neutral or charged current fermion absorption signals are inevitably unstable and their decays can be searched for using telescope observations. Since these indirect detection bounds are inherently model-dependent, we treat all of the dominant decays in concrete UV completions of the above dimension-6 fermionic absorption operators. For the neutral current, we present a model of gauged baryon-number with additional coupling to $ \chi $ and introduce a mixing of $ \chi $ with the (Dirac) neutrino. For the charged current, we present a modification of left-right symmetric models where $ \chi $ is put into a right handed doublet with the electron instead of the neutrino. In all cases, the decays depend on large powers of $m_\chi$ and so, requiring dark matter to be sufficiently long-lived leads us to consider masses well below the GeV scale.

This paper is organized as follows. For the neutral current and charged current dimension-6 operators which yield these unique signals, we present simple UV completions in Section II. Additionally, Section II contains a detailed discussion of the UV model-dependent dark matter decay modes that constrain our parameter space. In Sections III through V, we comprehensively consider all possible signals from absorption by nuclear targets at current and future direct detection, neutrino, and neutrino-less double beta decay experiments. We conclude in Section VI.

\newpage
\section{UV Completions}
\label{sec:UVcompl}
In this section, we present two UV completions that realize the neutral and charged current signals presented in this work. After presenting the models, we discuss in detail the various cosmological and collider constraints, with a particular emphasis on implications for dark matter stability. In general, a variety of thermal (and non-thermal) production mechanisms can accommodate the observed dark matter relic abundance. Therefore, accommodating $\Omega_\chi h^2 \sim 0.1$ does not place any restrictions on the model parameter space that is of interest to fermionic absorption signals. As such, we omit a detailed discussion of production mechanisms from the current work.

\subsection{Neutral Current}
\label{sec:NCUVcompl}
Simple models that generate the neutral current operator can be built through the introduction of additional $U(1)$ symmetries broken above the weak scale, and a mass-mixing between the dark matter candidate and a neutrino. For simplicity, consider a scenario where only $ \chi $ and the SM quarks are charged under a new $U(1)'$, with all of the quarks charged equally (i.e., gauging baryon number):
\begin{align} 
{\cal L}   \,\,\, \supset  \,\,\, g_\chi  \Big( \frac{1}{3} \, \sum _{ q }\bar{q}  \gamma _\mu q  +   Q_\chi \bar{\chi} \gamma _\mu \chi \Big) Z ^{ \prime \mu }  +  \frac{\epsilon}{2}  Z _{ \mu\nu } 'F ^{  \mu \nu  }+  \frac{m _{ Z ' } ^2 }{2} Z _\mu' Z ^{ \prime \mu } \,,
\label{eq:Zlag}
\end{align} 
where $ g_\chi  $ is the $U(1)'$ gauge coupling, we have taken the dark matter to have charge $Q_\chi$ under the $U(1)'$, and we have set the quark charge to unity without loss of generality. We also include a kinetic mixing, $ \epsilon $, which has a natural value $\epsilon \sim e g_\chi / 16 \pi ^2 $ arising from the running of quarks within the loop. Integrating out the $ Z ' $ yields the following dimension-6 operator:
\begin{align} 
\label{eq:IntZprimeout}
{\cal L}   \,\,\, \supset  \,\,\, \frac{ g_\chi ^2    }{ m _{ Z ' } ^2 }  \frac{1}{3} Q_\chi \sum _{ q } \bar{q}  \gamma^\mu q  \,\, \bar{\chi} \gamma _\mu \chi \,.
\end{align} 

Note that Eq.~\eqref{eq:IntZprimeout} is an operator typically considered in elastic scattering. Now suppose that $ \chi $ mixes with the SM neutrinos through a Yukawa interaction of a scalar, $\phi$ (with charge $Q_\chi$ under the $U(1)'$) which gains a vacuum expectation value (giving the $ Z ' $ a mass contribution). For simplicity, we consider a model with lepton number charged dark matter and approximately massless Dirac neutrinos such that the $U(1)'$ invariant mass term is given by:
\begin{align} 
{\cal L}_{\rm mass}  \,\,\, &\supset  \,\,\, m_\chi    \bar{\chi}  \chi   + \prn{y  \phi     \bar{\chi} P _R \nu  +{\rm h.c.}}    \,\,  =  \,\,
\big( \begin{array}{cc}\bar{\nu}  & \bar{\chi} \end{array} \big) \left( \begin{array}{cc} 
0   &0  \\  
y  \left\langle \phi \right\rangle   & m_\chi
\end{array} \right) P _R \left( \begin{array}{c} 
\nu  \\  
\chi  
\end{array} \right)  \,+ \,{\rm h.c.}  + \,...  
\end{align} 
After diagonalization, there is one massless state (identified with the SM neutrino) and one massive state with mass $ \sqrt{ m _\chi ^2 + y ^2  \left\langle \phi \right\rangle ^2} $.  Furthermore, a mixing is induced between $ \chi _R \equiv P_R \chi$ and $ \nu _R $ with a mixing angle, $ \theta _R $ given by:
\begin{equation} 
 s _{ \theta _R } = \frac{y \avg{\phi}}{\sqrt{y^2 \avg{\phi}^2+m_\chi^2}} \, .
\end{equation} 
Since the mixing is only between the right handed fields, the $ W $-induced $ \chi \rightarrow \nu \gamma  $ decay rate is heavily suppressed while maintaining a large direct detection signal (in contrast to the case of sterile neutrinos). 

We now discuss the phenomenology of this model. The direct detection signal is primarily governed by the effective operator: 
 \begin{equation} 
\label{eq:LNC}
{\cal L} \,\,\, \supset \,\,\, \frac{ Q_\chi g_\chi ^2   s _{ \theta _R } c _{ \theta _R }  }{ m _{ Z ' } ^2 }   \left( \bar{n}  \gamma _\mu n  + \bar{p} \gamma _\mu p \right)\bar{\chi}   \gamma _\mu P _R \nu   +{\rm h.c.} 
\end{equation} 
There will also be an elastic scattering mode but it is challenging to see for the masses of interest here, as it produces a smaller energy deposit. Consequently, searches looking for elastic scattering will generally be weaker than a dedicated fermion absorption search in this model. 

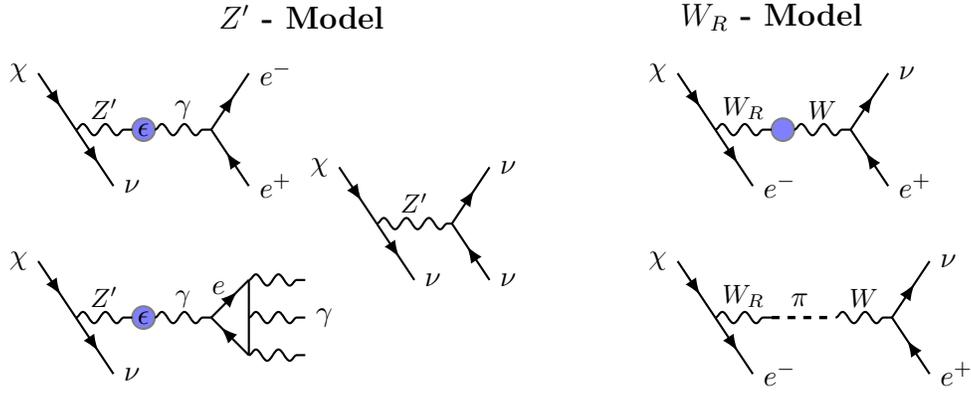
\begin{figure} 
\begin{center} \begin{tikzpicture} [line width=0.8]
    \node[] at (3.75,2.5) {\large \bf $Z' $ - Model};
    \node[] at (10,2.5) {\large \bf $W_R $ - Model};

    \draw[fb] (1.25,0.25) node[right] {$ \nu $} -- (0.75,1);
    \draw[f] (0.25,1.75)node[left] {$ \chi $} -- (0.75,1) ;
    \draw[v] (0.75,1) -- (1.5,1) node[above,midway] {\small$ Z ' $};
    \draw[fill=blue,opacity=0.5] (1.5+.15,1) circle (0.15);
    \node[] at (1.5+.15,1) {$\epsilon $};
    \begin{scope}[shift={(-0.25-0.75+0.3,0)}]
      \draw[v] (2.5,1) -- (3.25,1) node[above,midway] {$ \gamma $};
    \draw[f] (3.25,1) -- (3.75,1.75) node[right] {$ e ^- $};
    \draw[f] (3.75,0.25)node[right] {$ e ^+ $} -- (3.25,1) ;
  \end{scope}

  \begin{scope}[shift={(0,-2.5)}]
    \draw[fb] (1.25,0.25) node[right] {$ \nu $} -- (0.75,1);
    \draw[f] (0.25,1.75)node[left] {$ \chi $} -- (0.75,1) ;
    \draw[v] (0.75,1) -- (1.5,1) node[above,midway] {\small$ Z ' $};
    \draw[fill=blue,opacity=0.5] (1.5+.15,1) circle (0.15);
    \node[] at (1.5+.15,1) {$\epsilon $};
    \begin{scope}[shift={(-0.75+0.3,0)}]
      \draw[v] (2.25,1) -- (3.,1) node[above,midway] {\small$ \gamma $};

      \draw[decoration={markings,mark=at position .7 with {\arrow[very thick]{latex}}},postaction={decorate}] (3,1) -- (3.5,1.5);
      \draw[decoration={markings,mark=at position .75 with {\arrow[very thick]{latex}}},postaction={decorate}] (3.5,.5) -- (3,1);
      \draw[] (3.5,1.5) -- (3.5,.5);
      \draw[v] (3.5,.5) -- (4.25,.5);
      \draw[v] (3.5,1) -- (4.25,1) node[right] {$ \gamma $};
      \draw[v] (3.5,1.5) -- (4.25,1.5);
      \node[] at (3.1,1.4) {$e$};

\end{scope}

\end{scope}

\begin{scope}[shift={(4,-1.25)}]
    \draw[fb] (1.25,0.25) node[right] {$ \nu $} -- (0.75,1);
    \draw[f] (0.25,1.75)node[left] {$ \chi $} -- (0.75,1) ;
    \draw[v] (0.75,1) -- (1.75,1) node[above,midway] {\small$ Z ' $};
    \begin{scope}[shift={(-1.5,0)}]
    \draw[f] (3.25,1) -- (3.75,1.75) node[right] {$ \nu  $};
    \draw[f] (3.75,0.25)node[right] {$ \nu  $} -- (3.25,1) ;
  \end{scope}

\end{scope}

    \begin{scope}[shift={(8.5,4)}]
      \begin{scope}[shift={(0,-4)}]
    \draw[fb] (1.25,0.25) node[right] {$ e ^-  $} -- (0.75,1);
    \draw[f] (0.25,1.75)node[left] {$ \chi $} -- (0.75,1) ;
    \draw[v] (0.75,1) -- (1.5,1) node[above,midway] {\small$ W_R $};
    \draw[fill=blue,opacity=0.5] (1.5+.15,1) circle (0.15);
    \begin{scope}[shift={(-0.25-0.75+0.3,0)}]
      \draw[v] (2.5,1) -- (3.25,1) node[above,midway] {\small$ W $};
    \draw[f] (3.25,1) -- (3.75,1.75) node[right] {$ \nu  $};
    \draw[f] (3.75,0.25)node[right] {$ e ^+ $} -- (3.25,1) ;
  \end{scope}
\end{scope}

\begin{scope}[shift={(0,-6.5)}]
    \draw[fb] (1.25,0.25) node[right] {$ e ^-  $} -- (0.75,1);
    \draw[f] (0.25,1.75)node[left] {$ \chi $} -- (0.75,1) ;
    \draw[v] (0.75,1) -- (1.5,1) node[above,midway] {\small $ W_R $};
    \draw[snar] (1.5,1) -- (2.25,1) node[midway,above] {$ \pi $};
    \begin{scope}[shift={(-0.25-0.75+0.6+.25,0)}]
      \draw[v] (2.5,1) -- (3.25,1) node[above,midway] {\small$ W $};
    \draw[f] (3.25,1) -- (3.75,1.75) node[right] {$ \nu  $};
    \draw[f] (3.75,0.25)node[right] {$ e ^+ $} -- (3.25,1) ;
  \end{scope}
\end{scope}

\end{scope}
\end{tikzpicture}
\end{center}
\caption{Most constraining decays for $ \chi $ in the neutral current model ({\bf left}) and charged current model ({\bf right}).}
\label{fig:decays}\end{figure}
As alluded to above, dark matter is unstable as the $ \chi - \nu $ mixing can lead to various decays of $\chi$ depicted in  Fig.~\ref{fig:decays} ({\bf left}). Consider first those decays in Fig.~\ref{fig:decays} ({\bf left}) induced by 1-loop kinetic mixing without additional insertions of $ Z ' $ or $ Z $ propagators. A curious feature of this $ Z ' $ model is it \emph{does not} induce 1 photon or 2 photon decay channels up to these additional insertions --- the single photon channel through kinetic mixing is forbidden by gauge invariance (this is equivalent to the usual statement that particles charged under a new $U(1)'$ do not couple to the SM photon after diagonalization), while the $ 2 $ photon channel is forbidden by charge conjugation (also known as Furry's theorem). Considering higher orders, we find the decay of dark matter to 1, or 2 photons up to neutrino mass insertions:
\begin{align} 
\label{eq:NCdecays12photon}
&\Gamma _{ \chi \rightarrow \nu \gamma   }   \,\,  =  \,\, 0   +  {\cal O}   \left(  \frac{ m _\chi ^{13} }{ (4\pi)^{13} m _{ Z ' } ^{12}} \right)   , 
&\Gamma _{ \chi \rightarrow \nu \gamma \gamma    }  \,\,  =   \,\, 0   +   {\cal O}   \left(  \frac{ m _\chi ^9 }{ (4\pi)^{11} m _{ Z ' } ^8 } \right) .
\end{align} 
For the 1 photon channel, the dominant decay is through a 3-loop diagram with 3 $Z'$s. The leading contribution to the 2 photon channel comes from a 2-loop diagram with 2 $Z'$s. All of these contributions are negligible for the dark matter masses of interest to us here ($m_\chi \lesssim  m_\pi$). Including neutrino mass insertions induces decays through a $ W $ loop analogous to those of sterile neutrinos but suppressed by an additional mixing angle and dependent on the flavor structure between the right and left handed neutrinos. Since the neutrino that enters the effective operator in Eq.~\eqref{eq:LNC} via mixing with $\chi$ can be massless, these potential decay channels can be made arbitrarily small. We assume this here for simplicity.

For $m _\chi \gtrsim 2 m _e $ the dominantly constraining decay mode is $ \chi \rightarrow \nu e ^+ e ^- $  induced by kinetic mixing with decay rate given by
\begin{align} 
\label{eq:NCdecaysee}
\Gamma _{ \chi \rightarrow \nu e ^+ e ^-  } & \,\, = \,\, \left( 16 \log 2 - \frac{ 31 }{ 3} \right) \frac{ m _\chi ^5 }{ 512 \pi ^3 }   \left( \frac{ \epsilon e  Q_\chi g_\chi s _{ \theta _R }  c _{ \theta _R }}{  m _{ Z '} ^2 } \right) ^2    \,,
\end{align} 
For lower dark matter masses the dominant visible decay is $ \chi \rightarrow \nu \gamma \gamma \gamma $ through kinetic mixing in conjunction with the Euler-Heisenberg Lagrangian~\cite{Heisenberg:1935qt} (one can also circumvent kinetic mixing by attaching external photons to a loop of quarks however this diagram involves parametric suppressions by meson masses and we estimate it to be subdominant). The decay rate is estimated as (computing the phase space factor numerically with the aid of {\it MadGraph}~\cite{Alwall:2014hca}):
\begin{align} 
\label{eq:NCdecaysThreeGamma}
\Gamma _{ \chi \rightarrow \nu \gamma \gamma \gamma } & \,\, \simeq \,\,  10 ^{ - 7}  m _\chi ^{ 13}\left(   \frac{ ( 8 Q_\chi g_\chi \epsilon ) s _{ \theta _R } c _{ \theta _R } \alpha ^2 }{ 360 m _e ^4 m _{ Z ' } ^2 } \right) ^2 \,.
\end{align} 
In addition, $ \chi $ can decay invisibly to neutrinos, $ \chi \rightarrow 3 \nu $, which proceeds through a large power of the $\nu_R - \chi_R$ mixing angle:  
\begin{align} 
\label{eq:NCdecaysThreeNu}
\Gamma _{ \chi \rightarrow \nu \nu \nu } & \,\,=\,\, \left( 16 \log 2 - 11\right) \frac{m _\chi ^5}{ 128 \pi ^3 } \left( \frac{ Q_\chi^2 g_\chi ^2 s _{ \theta _R } ^3 c _{ \theta _R }}{ m _{ Z ' } ^2} \right) ^2  \,.
\end{align} 
All the decays arise from irrelevant operators, and as such the rates are proportional to large powers of $m_\chi$. Therefore, ensuring a stable dark matter candidate leads us to consider lighter dark matter candidates. The limits on dark matter decay rates depends sensitively on the dark matter mass and particular decay channel. For $ \chi \rightarrow \nu e ^+ e ^- $ and $ \chi \rightarrow \nu \gamma \gamma \gamma $ decays we recast constraints from~\cite{Essig:2013goa} while for $ \chi \rightarrow  \nu \nu \nu$ we use bounds from the non-observation of an anomalous change in the equation of state of the Universe from the era of the Cosmic Microwave Background until present day~\cite{Gong:2008gi}. 

The particular decay rates computed here clearly depend sensitively on the particular model chosen. As a striking example, note that in the case of a scalar mediator its possible to completely eliminate the $ \chi \rightarrow 3 \nu $ decay mode by choosing a scalar which does not carry a coupling to two neutrinos. In an effort to not let the specifics of the model overshadow the signal regions observable in experiments, we allow for the possibility of fine-tuning away decays by introducing a UV kinetic mixing parameter and a UV contribution to the $ 3 \nu $ operator that can cancel these decays modes to some level. As we show, this will be necessary in all the detectable parameter space of neutral current absorption for $ m _\chi \gtrsim ~{\rm MeV} $. 

Another possible tension could be that the production of dark matter results in too great an energy density in the right-handed neutrinos. Assuming that dark matter is produced via UV freeze-in, we find that for the lightest dark matter masses we consider for the NC operators, the energy density in $\nu_R$ relative to that in a SM $\nu_L$ is always less than $\sim 10^{-3}$. Thus, though the energy density in $\nu_R$ depends on the initial DM production mechanism, in general, it does not have to be in tension with measurements of the early radiation energy density.

In addition to constraints from indirect detection, there are bounds on this UV completion that do not depend on $\chi$ being dark matter. For $ m_{Z'}$ well above the weak scale, the dominant constraints arise from mono-jet searches (see~\cite{Belyaev:2018pqr} for a recent summary). For lighter $ Z ' $, the dominant constraints arise from flavor changing meson and $Z$ decays induced by a Wess-Zumino-Witten term present in theories which gauge an anomalous combination of SM charges~\cite{Dror:2017ehi,Dror:2017nsg}. Constraints also come from looking for heavy anomaly-canceling fermions directly in colliders~\cite{Dobrescu:2014fca}. Since the scale of the effective operators we consider here are above the weak scale, we do not expect significant constraints from star cooling, beam dump, or supernovae which are often crucial when discussing light dark matter.~\footnote{In principle, this conclusion may be too hasty since, while the fermion absorption operator scale we consider will always be above the weak scale, the $ \bar{q} \gamma ^\mu q \bar{\chi} \gamma _\mu \chi $ operator could have a scale a little below the weak scale. Nevertheless, since we work in a regime where it is at most comparable to the weak scale and $ Z ' $ only couples to baryons we estimate there are no additional strong constraints.} 

\subsection{Charged Current}
\label{subsec:CCmodel}
UV completions which result in a charged current signal typically require new states charged under electromagnetism. Such a situation is a prediction of an extended electroweak sector, one example of which we explore here. A simple extended breaking pattern is~\cite{Senjanovic:1975rk}\footnote{More generally, the fermions may be charged under SU(2)$ _{L/R} $ in alternative structures (see e.g. \cite{Hsieh:2010zr} for a review).}
\begin{align} 
{\rm SU(2)}_  {L} \times  {\rm SU(2)}_R \times {\rm U(1)}_X  \xrightarrow{\left\langle \Phi \right\rangle} {\rm SU(2)}_  L \times {\rm U(1)}_Y  \xrightarrow{\left\langle H \right\rangle} {\rm U(1)}_{ \rm EM}  \,.
\end{align} 
The initial breaking can be accomplished when an $SU(2)_R $ doublet scalar, $ \Phi $ charged as $ ( \mathbf{1},\mathbf{2},1/2 ) $, gets a vev;
\begin{equation} 
\Phi = \left( \begin{array}{c} 
\phi ^+  \\  
\phi ^0  
\end{array} \right) \,, \quad \left\langle \Phi \right\rangle = \frac{1}{\sqrt{2}} \left( \begin{array}{c}  
0 \\  
u 
\end{array} \right)  \,.
\end{equation} 
In this stage of breaking $U(1) _Y $ is formed out of a linear combination of $SU(2)_R \times  U(1)_X  $ charges:
\begin{equation} 
Y = X + T _R ^3   \,.
\end{equation} 
The second stage of breaking can be accomplished with a $ H $ charged as $  ( \mathbf{2},\bar{\mathbf{2}} ,0 ) $. This corresponds to:
\begin{equation} 
H = \left( \begin{array}{cc} 
h _1 ^0  & h_1 ^+  \\  
h _2 ^-  & h _2 ^0 
\end{array} \right) \,, \quad \left\langle H \right\rangle = \frac{ v }{ \sqrt{2} } \left( \begin{array}{cc} 
c _\beta  & 0 \\  
0 & s _\beta 
\end{array} \right)  \,,
\end{equation} 
where the EM charge is given by,
\begin{equation} 
Q = X + T _L ^3 + T _R ^3  \,.
\end{equation} 

The lepton number carrying dark matter $\chi$ in this set-up is identified with the right handed component $\chi_R \equiv P_R \chi$ charged as $(\bold{1}, \bold{2}, 0)$ and is assumed to complete the lepton right handed doublets.  We do not need to introduce gauge singlet right-handed neutrino partners for the SM neutrinos, but may do so to realize a standard seesaw mechanism. Many known mechanisms may be used to generate SM neutrino masses and we do not prefer a particular one as they do not affect the fermionic absorption phenomenology. Additionally, there will be an inert left handed component $\chi_L$ \emph{i.e.}, a singlet under all gauge symmetries. As is typical for left right symmetric models we place the SM right-handed fermions (we will consider only one generation here) into right handed doublets ($ \equiv R  $) under $SU(2)_R $ and left handed fermions in doublets ($ \equiv L  $) under $SU(2)_L $. 

The $SU(2) _R $ gauge boson masses primarily arise in the usual way, from the kinetic term once $\Phi$ develops a vev $ \left\langle \Phi \right\rangle $, and are given by
\begin{equation} 
M _{W _R } = \frac{1}{2} g _R u \,, \quad  M _{ Z _R } = \frac{1}{2} ( g _R ^2 + g_\chi ^2 ) ^{1/2} u \,.
\end{equation} 
In addition, there is a mass mixing between the $ W $ and $ W_R $ at tree level given by,
\begin{equation} 
{\cal L}  \,\,\,  \supset   \,\,\,  - \frac{1}{4}g _L g _R v ^2 s _{ 2 \beta  }  W _\mu W_R^\mu \,.
\end{equation} 
For a generic scalar potential $ s _{ 2 \beta } $ is $ {\cal O} ( 1  ) $ and hence there is a mixing angle between $W $ and $ W _R $ of ${\cal O} (  m _W ^2  / m _{ W _R } ^2 ) $, which leads to $ \chi $ decay. To minimize this mixing we work in limit that $ s _{ 2 \beta } \rightarrow 0 $, which can be achieved if $ H $ contributes negligibly  to the breaking of $SU(2) _R $ such that $ c _\beta = 1 $ and $ s _\beta = 0 $ as in the inert doublet model~\cite{Deshpande:1977rw}. 

Recall that we place the quarks into right handed multiplets, while the dark matter $\chi_R$ completes the lepton right handed doublets. This leads to the following term allowed by all the symmetries: 
\begin{align} 
{\cal L}  \,\,\,  \supset  \,\,\,  \frac{g_R}{\sqrt{2}} W_{R \mu} \prn{\bar{\chi} \gamma^\mu P_R e +  \bar{u} \gamma^\mu P_R d} + \text{h.c.} \,.
\end{align} 
Additionally, fermion masses are generated from Yukawa interactions with $ H $ and $ \tilde{H} \equiv \sigma_2 H ^\ast \sigma_2  $ of the form $\bar{L} H R$ and $\bar{L}  \tilde{H}  R$ as follows:
\begin{align} 
{\cal L}  \,\,\, \supset \,\,\, \frac{ y _u  v }{ \sqrt{2} } \bar{u}   u  + \frac{ y _d  v }{ \sqrt{2} } \bar{d}  d  + \frac{ y _{\ell}  v }{ \sqrt{2} } \bar{\ell}  \ell  + \frac{ y _\nu  v }{ \sqrt{2} } \prn{\bar{\nu}P_R \chi + \text{h.c.}}\,.
\end{align} 

Unlike standard studies, instead of considering right handed neutrinos in the lepton doublets we have introduced $ \chi _R   $ states as well as the additional inert $ \chi _L  $. In this sense this model explicitly breaks the true left-right symmetric nature of the setup. Since $ \chi _L $ is a singlet it forms a Yukawa coupling with the $ \Phi $ and the right handed doublet,
\begin{align} 
{\cal L} &  \,\,\,  \supset  \,\,\,  y _\chi \, \Phi{\bar \chi}  \, P _R  \left( \begin{array}{c} 
e   \\  
 \chi  
\end{array} \right) +{\rm h.c.} \quad \supset \quad  \frac{ y_\chi \, u }{ \sqrt{2} } \bar{\chi} \chi ,
\end{align} 
preventing $ \nu _L $ and $ \chi _R $ from forming a Dirac fermion. We assume $y_\chi u \gg y_\nu v$ such that the SM neutrino is effectively massless and $m_\chi$ is a free parameter. \footnote{$y_\nu$ is not needed, but is included since no symmetry forbids it. However, we do assume that it is sufficiently small to prevent significant decays to 3 left-handed neutrinos through the SM 4-neutrino coupling. }
After integrating out the $ W_R $ boson we get the quark level interactions:
\begin{equation} 
{\cal L} \,\,\, \supset \,\,\, \frac{ g _R ^2 }{ M _{ W _R } ^2 } \frac{1}{2} \bigl[ \bar{u} \gamma _\mu P _R d  \bigr] \bigl[  \bar{e} \gamma ^\mu P _R \chi  \bigr] + \text{h.c.} \,.
\end{equation} 
In terms of the nucleons this gives the interaction:
\begin{equation} 
\label{eq:CClagrangian}
{\cal L} \,\,\, \supset \,\,\, \frac{ g _R ^2 }{ 4M _{ W _R } ^2 } \bigl[ \bar{p} \gamma _\mu \left( 1  +  \lambda    \gamma _5  \right)  n  \bigr] \bigl[  \bar{e} \gamma ^\mu P _R \chi  \bigr] + \text{h.c.} \,,
\end{equation} 
where $  \lambda   \simeq  1.2694 \pm 0.0028$ is the axial to vector coupling from data~\cite{Formaggio:2013kya}.

We now consider possible decays as shown in Fig.~\ref{fig:decays} ({\bf right}). The safest possibility is to only charge the first generation under the new $SU(2)_R$ to minimize the mixing between the $W$ and $W_R$, and so we focus on this case\footnote{In addition to making the $W$-$W_R$ mixing worse, charging more SM generations would introduce new, accompanying dark-sector states for the heavier SM leptons. If they comprise a fraction of dark matter, they would not produce charged current signals at experiments as processes with a dark matter absorbed and a heavy lepton emitted would be kinematically forbidden. Thus, the charged current signals would only come from the fraction of ``first generation'' dark matter.}. One loop radiative corrections induce a log-divergent mixing between $ W _R $ and the SM $ W $ boson which vanishes at $u $, and at low energies, is approximately
\begin{equation} 
{\cal L} \,\,\, \supset \,\,\, \left( \frac{ g _L  ^2 g _R ^2  m _u m _ d  }{ (4\pi)^2 M _{ W_R } ^2 M _W ^2 } \log \frac{ u }{ \Lambda_{\rm QCD} } \right) g _{\mu\nu} j _L ^\mu j _{R }^{  \nu } 
\end{equation} 
where $ j _{ L,R} ^\mu $ are the left and right gauge currents (defined without the couplings). Below the QCD scale there is an additional contribution from the running which we estimate at leading order using chiral perturbation theory. Starting with the chiral Lagrangian ($ \Sigma \equiv e ^{ i \pi ^a \sigma ^a / f_\pi }$) we can extract the mixing with the pions:
\begin{equation} 
\frac{ f _\pi ^2 }{ 4 } {\rm Tr} \left[ ( D _\mu \Sigma ) ^\dagger D ^\mu \Sigma \right] \,\, \supset \,\,  \frac{ f _\pi }{ 2 } \left[ g _L W ^+ _\mu  + g _R W _{R\, \mu } ^+ \right ] \partial ^\mu \pi ^-  +{\rm h.c.} 
\end{equation} 
Integrating out the pions induces a coupling between the left and right handed currents:
\begin{equation} 
{\cal L}  \,\,\, \supset  \,\,\,  \left(  \frac{ g _L ^2 g _R ^2 f _\pi ^2 }{ 8m _\pi ^2 }  \frac{1}{ M _{W _R } ^2 M _W ^2 } \right) \partial ^\mu j _{R, \mu } \partial ^\nu j _{ L , \nu }  +{\rm h.c.}\,,
\end{equation} 
which of the two terms dominates will depend on the mass of $ \chi $. The decay rates (ignoring the interference terms) are:
\begin{align} 
\label{eq:CCdecayrates}
\Gamma _{ \chi \rightarrow e ^+  e ^- \nu  } ^{ ( {\rm 1-loop} )} & = \left( 16 \log 2 - 11 \right) \frac{m _\chi ^5}{ 512 \pi ^3  }  \left( \frac{ g _L  ^2 g _R ^2  m _u m _ d  }{ (4\pi)^2 M _{ W_R } ^2 M _W ^2 } \log \frac{ m_u }{ \Lambda_{\rm QCD} } \right) ^2    ,\\ 
\Gamma _{ \chi \rightarrow e ^+ e ^- \nu } ^{ ( \pi ) }& = \left( \log 4 -\frac{31}{24} \right) \frac{m _\chi ^{7} m _e ^2 }{ 256 \pi ^3  }  \left(  \frac{ g _L ^2 g _R ^2 f _\pi ^2 }{ 8m _\pi ^2 }  \frac{1}{ M _{W _R } ^2 M _W ^2 } \right) ^2 .
\end{align} 
Note that Eq.~\eqref{eq:CCdecayrates} will in general be less constraining than decays in the neutral current model as Eq.~\eqref{eq:CCdecayrates} contains additional factors of inverse mediator when compared to Eq.\eqref{eq:NCdecaysee}- \eqref{eq:NCdecaysThreeNu}. 

As in the case of the neutral current, one can look directly for the operators we consider here without requiring the presence of $ \chi $ as dark matter. The most powerful direct search arise from collider physics from searches for heavy charged states. For simplicity we focus on the limit where $ W _R $ is heavy such that it is never produced on-shell. In this case, there are limits using the energy-enhanced nature of the $ u \bar{d} \rightarrow e \chi $ process however we note that this process does not interfere with any Standard Model rate resulting in most collider searches being inapplicable (as they rely on a final state neutrino). Nevertheless, there are searches at 8 TeV which look for helicity-non-conserving contact interactions which should be roughly applicable here~\cite{Khachatryan:2014tva}. We estimate that these restrict $ g _R ^2 / 4 M _{W _R } ^2 \lesssim ( 4.5 ~{\rm TeV} ) ^{ - 2 } $. 

In addition to direct collider searches, one can look for deviations from the SM in known $\beta$ decays. For any such decay, the SM prediction is hard to evaluate rendering it challenging to use these process for precision searches for new physics in the limit that the scale of the higher dimensional operator is well above the weak scale. Nevertheless, it was suggested to use super-allowed (Fermi) transitions in between isotopes with vanishing spin and unit parity ($ I ^P =  0 ^+ \rightarrow 0 ^+ $)~\cite{Hardy:2014qxa} (see also~\cite{Towner:1973yrc,Hardy:1975eq,Hardy:1990sz,Hardy:2004id,Hardy:2004dm,Hardy:2008gy} for earlier work). Such transitions are insensitive to the axial part of the operator and the vector contribution does not get renormalized under QCD~\cite{PhysRev.109.193} (what became known as the {\em conserved vector current} hypothesis), which makes it possible to compute the rates to the sub-percent level. In~\cite{Hardy:2014qxa}, constraints are put on operators of the form, $   \Lambda _\nu ^{ - 2 }\left[ \bar{p} \Gamma _i n \right] \left[ \bar{e} \Gamma _j  \nu \right] $, which can interfere with the SM amplitudes resulting in a limit on the operator cutoff scale: $ \Lambda _\nu  ^2 \gtrsim  10 ^{ 3} \, G _F $ ($ G _F \simeq 10 ^{ - 5} ~{\rm GeV} ^{ - 2 } $ is the Fermi constant). Computing these constraints for the operators of interest here is an involved task and beyond the scope of this work. Instead, we roughly estimate the sensitivity $ \beta $ decay experiments can have assuming a similar analysis can be done for operators involving $ \chi $. The constraint on $ \Lambda _\nu $ is sensitive to the interference term between the SM and new operator term which is $ {\cal O} ( G _F \, \Lambda _\nu  ^{-2} ) $, while for $ \chi $ operators there is no interference. Defining the higher dimensional operator for charged current fermion absorption with a scale $ \Lambda $, the leading term in the $ \beta $ experiments is $ {\cal O} \left( \Lambda ^{ - 4 } \right) $. Equating the observed limit to this operator we find that if such a search were carried out we would expect a sensitivity of order, $ \Lambda \gtrsim 1.5 ~{\rm TeV}$, which is weaker than present collider bounds. We also emphasize that such constraints depend critically on the mass of $ \chi $ --- when $ m _\chi \gtrsim \mathcal{O}\left( \text{MeV} \right)$ different $ \beta $ decay channels become kinematically unavailable, quickly weakening the constraints. Other possible ways to handle the nuclear uncertainties are using the neutron lifetime and angular distributions in nuclear decays, however the constraints using these techniques are weaker than the ones estimated above~\cite{Gonzalez-Alonso:2018omy}.

Outside of nuclear decays its possible to use charged pions decay searches looking for $ \pi ^\pm \rightarrow e ^\pm \chi  $~\cite{Aguilar-Arevalo:2017vlf}, however these searches are not able to extend to $ \chi $ masses below 60 MeV due to backgrounds from muon decays, which will be outside our range of interest for the charged current operator. Lastly, we comment that, as for the neutral current operator UV completion, we do not expect significant constraints from star cooling, beam dump, or supernovae since the scale of the effective operators we consider here are above the weak scale.

\section{Neutral Current Nuclear Recoils}
\label{sec:neutral}
We first study the nuclear recoils from the dimension-6 neutral current operator generated by the UV model discussed in Sec.~\ref{sec:NCUVcompl} with the identification $1/\Lambda ^2 \equiv Q_\chi g_\chi  ^2 s _{ \theta _R } c _{ \theta _R } / m _{ Z ' } ^2$,
\begin{equation} 
\label{eq:NCops}
\frac{1}{\Lambda^2}\prn{ \bar{n} \gamma ^\mu n + \bar{p} \gamma ^\mu p} \bar{\chi} \gamma _\mu P _R \nu   + \text{h.c.}\,.
\end{equation} 
This operator leads to the the nuclear recoil process;
\begin{align} 
\chi (m_\chi \vec{v}) + \text{N} (\vec{0}) \to \nu (\vec{p}_\nu) +  \text{N} (\vec{q}) \, ,
\end{align} 
in which an incoming dark matter with velocity $\vec{v}$ is absorbed by a target nucleus N at rest which then recoils with momentum $\vec{q}$ against the light $\nu$ of momentum $\vec{p}_\nu$. While this vector structure is inspired by the $Z'$ model, we emphasize that these signals can arise from operators with a more general Lorentz structure for which the formalism that follows can also be applied. 

The relevant experimental observable is the differential scattering rate per nuclear recoil energy. We begin with the usual differential cross section,
\begin{align} 
\label{eq:NCdsigma}
d \, \sigma & \, = \,  \frac{ \overline{\magn{\MM_N}^2}  }{4 E_\chi E_N v } \prod_j \frac{d^3p_j}{2E_j \prn{2\pi}^3}\prn{2\pi}^4  \delta^4 \prn{p_\chi^\mu + p_i^\mu-p_\nu^\mu-p_f^\mu}  \\
&\, = \,  \sqrt{\frac{E_R}{2M}}\frac{\overline{\magn{\MM_N}^2} dE_R \, d\prn{  \cos \theta_{qv}}}{16 \pi v m_\chi p_\nu}  \delta \left( E_R+p_\nu-m_\chi ( 1+v^2/2 ) \right) , \nonumber
\end{align} 
where $\overline{\magn{\MM_N}^2} $ is the matrix element squared averaged over initial and summed over final spins, $p_{i (f)}^\mu$ is the initial (final) four-momentum of the nucleus, $E_R=q^2/2M$ is the energy of the recoiling nucleus, $M$ is the mass of the nucleus, $\theta_{qv}$ is the angle between $\vec{v}$ and $\vec{q}$, and $p_\nu = \sqrt{m_\chi^2 v^2 +q^2 - 2m_\chi v q \cos \theta_{qv}}$. The incoming dark matter is non-relativistic, so its energy is roughly equal to its mass. Dropping $\ord{v}$ terms, the energy-conserving $\delta$ function simplifies to\footnote{$\nu_R$ could in fact be any neutral, light fermion in the dark sector. As long as it is much lighter than dark matter, $q\sim m_\chi \gg m_{\nu_R}$ gives rise to the distinctive fermionic absorption nuclear recoil spectrum.}
\begin{align} 
\label{eq:NCdeltaatv0}
\delta \left( E_R+p_\nu-m_\chi ( 1+v^2/2 ) \right) \simeq  \frac{m_\chi}{M} \delta \prn{E_R-E_R^0}, 
\end{align} 
where $E_R^0=m_\chi^2/2M$ since $m_\chi \ll M$. Thus, the differential cross section reduces to
\al{
\frac{d \sigma}{d E_R}=\frac{\overline{\magn{\MM_N}^2}}{16 \pi v M^2} \delta \prn{E_R-E_R^0}.
}

The differential scattering rate per nuclear recoil energy in an experiment is related to this differential cross section by
\al{
\label{eq:NCdRdERdefn}
\frac{ d R }{ d E _R } = N_T n_\chi \avg{\frac{d \sigma}{d E_R} v}  \Theta ( E _R ^0 - E _{ {\rm th}} ),
}
where $N_T$ is the number of nuclear targets in the experiment, $n_\chi$ is the local number density of dark matter, the average is performed over the incoming dark matter's velocity distribution, and $ \Theta ( E _R ^0 - E _{ {\rm th}} )$ approximates the nuclear recoil energy threshold of the experiment with a step function (see Appendix~\ref{sec:appExp} for a summary of $E _{ {\rm th}}$ for the experiments considered here). The average over the dark matter velocity distribution 	is trivial and yields
\begin{align} 
\label{eq:NCdRdERatv0}
 \frac{ d R }{ d E _R }  =N_T \frac{\rho_\chi}{m_\chi} \sigma_{\text{NC}} A^2 F ( q )  ^2 \delta ( E _R - E _R ^0 )  \Theta ( E _R ^0 - E _{ {\rm th}} )\,, 
\end{align} 
where $\rm \rho_\chi \simeq 0.4 \, GeV/cm^3$ is the local dark matter energy density, $\sigma_{\text{NC}}=m_\chi^2/\prn{4\pi \Lambda^4}$ is the absorption cross section per nucleon, $A$ is the atomic mass number of the target nucleus, and $F(q)$ is the Helm form factor~\cite{Lewin:1995rx} of the target nucleus (normalized to 1). 

This scattering rate is different from the usual elastic scattering rate for dark matter (see~\cite{Lin:2019uvt} for a recent review) since the typical recoil energy from an elastic scatter is of the order $v^2 \mu_{\chi N}^2/M$ (where $ \mu  _{ \chi N } $ is the $ \chi$-$N $ reduced mass), while the fermionic absorption recoil is peaked at $m_\chi^2/2M$. Therefore, for a fixed dark matter mass, the nuclear recoil energy for fermionic absorption is $ 1/ v ^2 \sim 10 ^{ 6 } $ times larger than that of usual elastic recoil. This allows direct detection experiments to probe dark matter candidates roughly $1/v \sim 10^3$ times lighter than normal, in addition to allowing neutrino detectors with larger exposures but higher thresholds to make competitive searches. 
\begin{figure}[t!]
\centering
\includegraphics[width =0.78\textwidth]{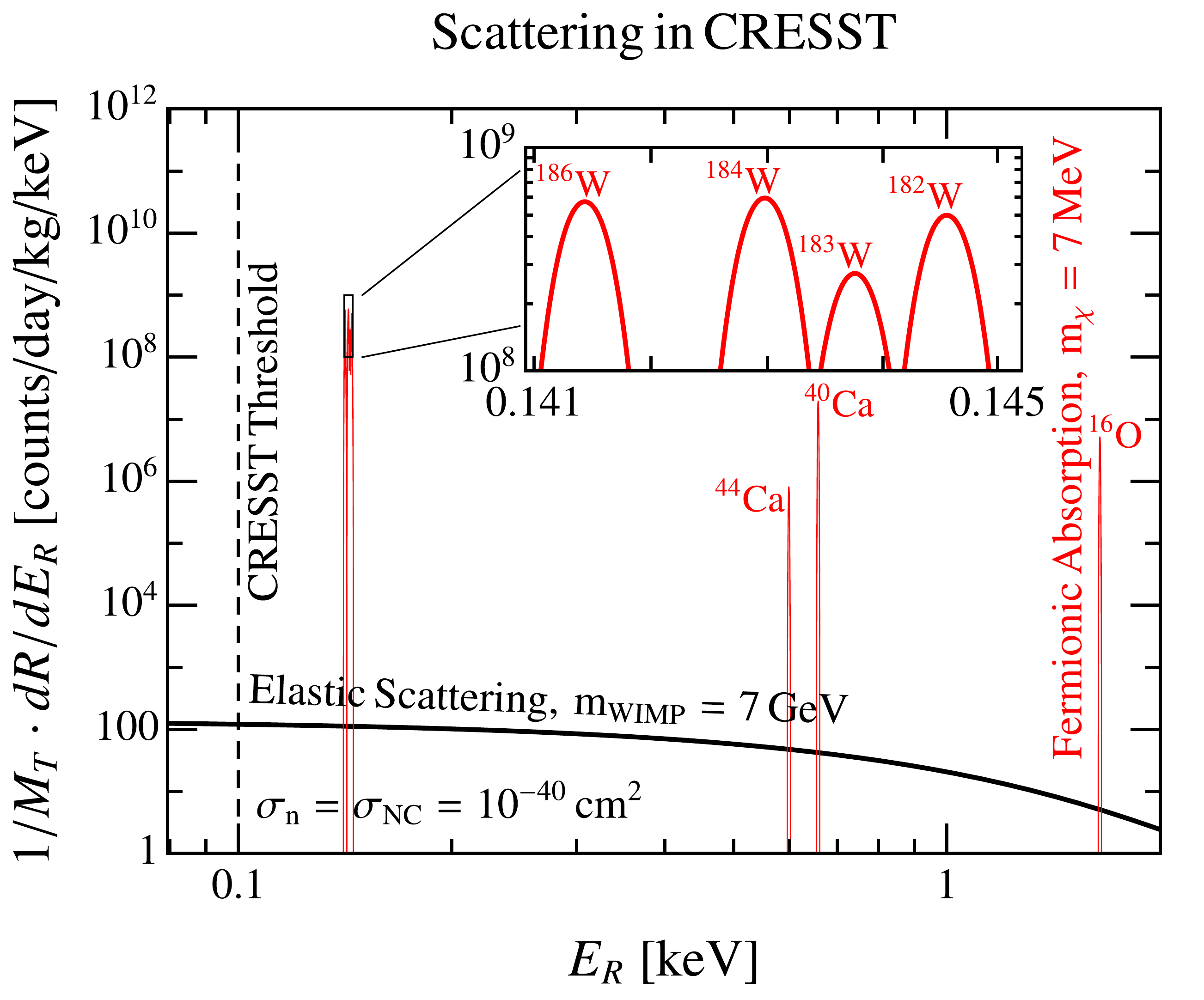}
\caption{\label{fig:dRdERCRESST} 
Differential scattering rate per recoil energy per detector mass at CRESST~\cite{Petricca:2017zdp} from fermionic absorption of a $m_\chi = 7 \text{ MeV}$ dark matter with $\sigma_{\text{NC}}=10^{-40} \text{ cm}^2$. Also shown is the elastic scattering rate for a WIMP with mass $m_{\text{WIMP}} = 7 \text{ GeV}$ and spin-independent cross section per nucleon $\sigma_n = 10^{-40} \text{ cm}^2$, along with the CRESSTIII nuclear recoil threshold at 100 eV. The figure inset zooms in on the bunched peaks corresponding to the four isotopes of Tungsten.}
\end{figure}
In order to highlight the differences between fermionic absorption and elastic scattering rates, we compare the differential scattering rates per recoil energy per detector mass ($M_T$) at one particular experiment, CRESST~\cite{Petricca:2017zdp}, in Fig.~\ref{fig:dRdERCRESST}. To make an illustrative comparison, we set the spin-independent WIMP cross-section equal to the absorption cross section per nucleon, which we set as $ \sigma _{\rm NC} =10^{-40} \text{ cm}^2$, and show the elastic rate for a heavier WIMP, $m_{\text{WIMP}}= 7 ~{\rm GeV} $, while taking $ 7 ~{\rm MeV} $ for the fermion absorption signal. To obtain the finite heights and widths of the fermionic absorption peaks which are not given by the $\delta$ function in Eq.~\eqref{eq:NCdRdERatv0}, we calculate the differential scattering rate after expanding the energy-conserving $\delta$ function to first order in $v$ (see Appendix~\ref{sec:NCdEdERderiv} for details). CRESST illustrates the differences in scattering rates well because it contains multiple target isotopes in its $\rm CaWO_4$ crystals which give rise to four peaks from absorbing fermionic dark matter which are distinguishable if the energy resolution is less than 50 eV~\cite{Petricca:2017zdp}. The figure demonstrates the relative ease with which experiments looking for fermionic absorption nuclear recoils can see the signal above the background by correlating the locations and heights of scattering rates off multiple target isotopes. Even in the absence of multiple distinguishable peaks, detectors can still use the peaked nature of the fermionic absorption differential scattering rate to differentiate the signal from the noise.

Having discussed the novel signature of neutral current nuclear recoils from fermionic absorption, we now project the sensitivities of future and current experiments to this signal. Integrating the differential scattering rate over all recoil energies and summing over all isotopes $j$ present in an experiment, we find the total event rate is
\begin{figure}[t!]
\centering 
\hspace{-0.9cm}
\includegraphics[width =0.78\textwidth]{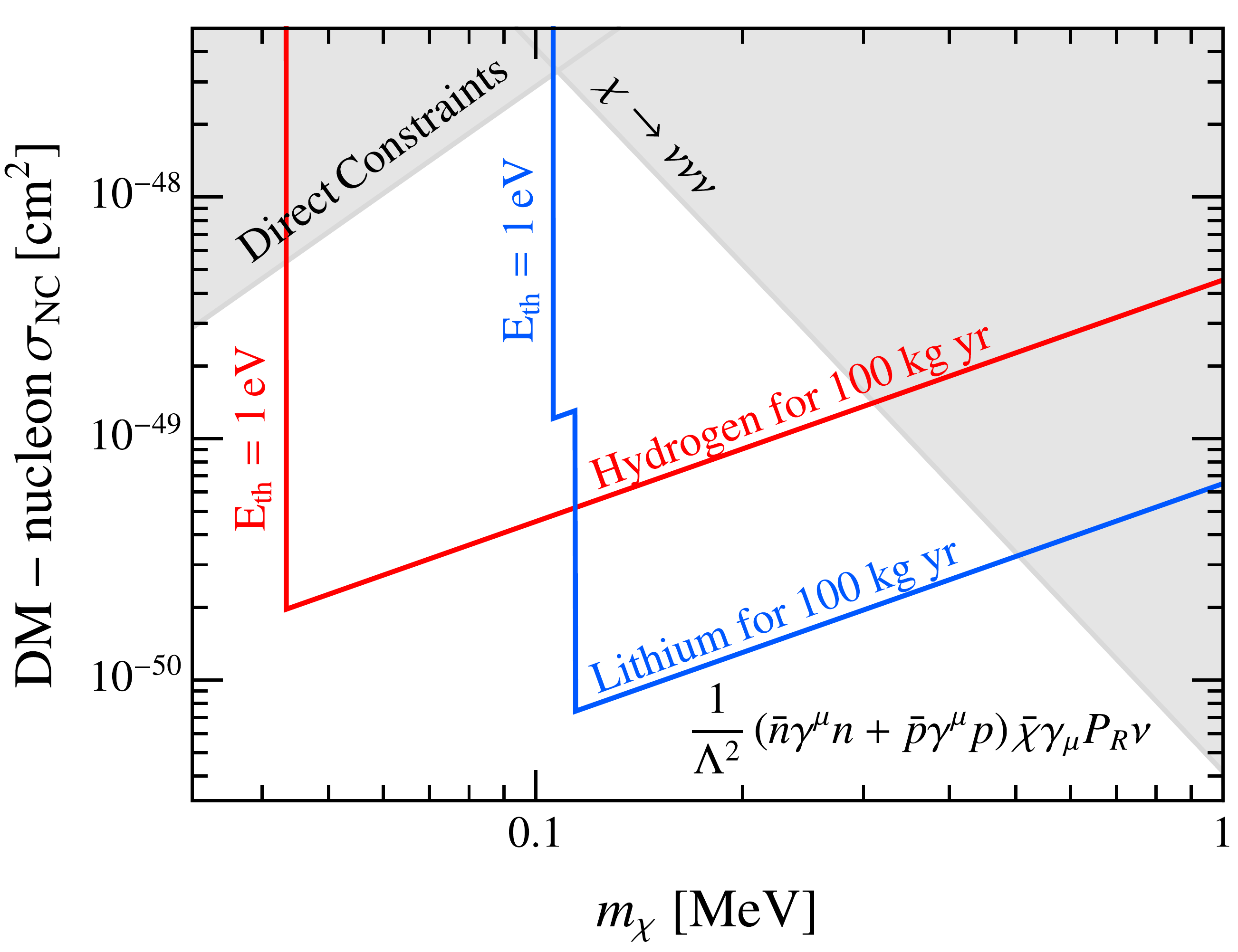}
\caption{Projected upper bound on $\sigma_{\text{NC}}$ as a function of $m_\chi$ at future detectors with Hydrogen or Lithium targets. Bounds for both potential targets are shown assuming $E_{\rm th} = 1 \text{ eV}$ and $M_T T = 100 \text{ kg yr}$. Also shown in gray are the constraints from direct searches for $ Z '$s and decays of $ \chi $ for the benchmark UV completion with $ m _{ Z '  } = 18~ {\rm GeV} $, $ s _{ \theta _R } = 10 ^{ - 1.5} $, and $ Q _\chi  = 0.1 $, as described in the text.} 
\label{fig:NCsigmalight} 
\end{figure}
\begin{align} 
\label{eq:NCrate}
R = \frac{\rho_\chi}{m_\chi} \sigma_{\text{NC}} \sum_j N_{T, j} A_j^2 F_j (q) ^2  \Theta ( E _{R, j} ^0 - E _{ {\rm th}} ) \,. 
\end{align} 
For simplicity, we project bounds on $\sigma_{\text{NC}}$ by requiring $< 10$ events occur in a given experiment. 

\begin{figure*}[t!]
\centering
\hspace{-0.9cm}
\includegraphics[width = 0.9 \textwidth]{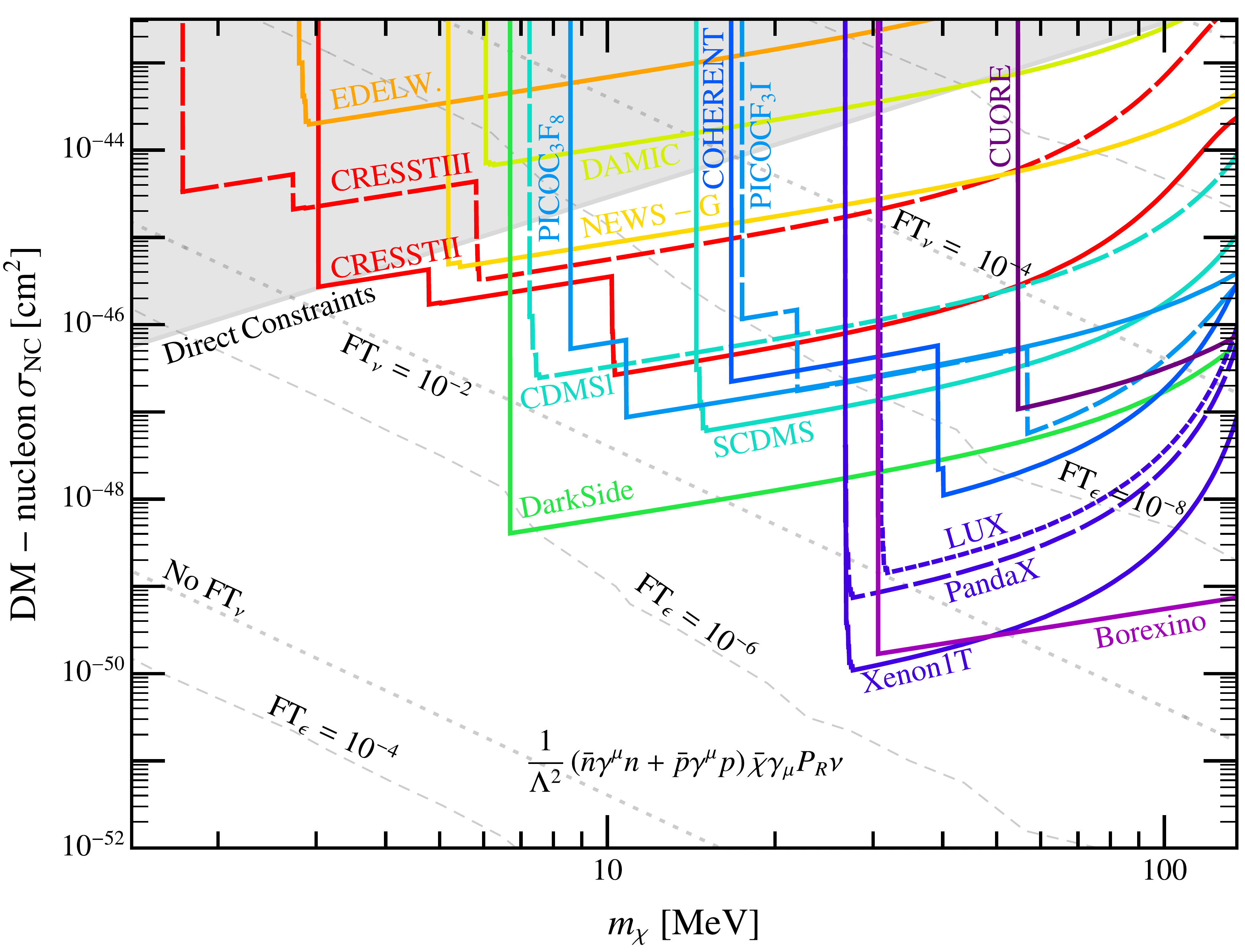}
\caption{\label{fig:NCsigma} 
Projected upper bound on $\sigma_{\text{NC}}$ as a function of $m_\chi$ at current experiments, including CUORE~\cite{Alduino:2017ehq}  (\textcolor{mydrkpurp}{dark purple}), Borexino~\cite{Agostini:2018fnx} (\textcolor{mypurple}{purple}), LUX~\cite{Akerib:2016vxi} (dotted \textcolor{mynavy}{navy blue}), PandaX-II~\cite{Cui:2017nnn} (dashed \textcolor{mynavy}{navy blue}), XENON1T~\cite{Aprile:2018dbl} (solid \textcolor{mynavy}{navy blue}), COHERENT~\cite{Akimov:2017ade,Scholberg:2018vwg} (\textcolor{myblue}{blue}), PICO-60 run with $\rm CF_3I$~\cite{Amole:2015pla} (dashed \textcolor{mysky}{sky blue}), PICO-60 run with $\rm C_3F_8$~\cite{Amole:2017dex} (solid \textcolor{mysky}{sky blue}), SuperCDMS~\cite{Agnese:2014aze} (solid \textcolor{myaqua}{aqua}), CDMSlite Run 2~\cite{Agnese:2015nto} (dashed \textcolor{myaqua}{aqua}), DarkSide-50~\cite{Agnes:2014bvk,Agnes:2018ves} (\textcolor{mygreen}{green}), 
 DAMIC~\cite{Aguilar-Arevalo:2016ndq} (\textcolor{mylime}{lime}), NEWS-G~\cite{Arnaud:2017bjh} (\textcolor{myyellow}{yellow}), EDELWEISS-SURF~\cite{Armengaud:2019kfj} (\textcolor{myorange}{orange}), CRESST II~\cite{Angloher:2015ewa} (solid \textcolor{myred}{red}), and CRESST III~\cite{Petricca:2017zdp} (dashed \textcolor{myred}{red}). Here we have taken $ m _{ Z '  } = 18 ~{\rm GeV} $, $ s _{ \theta _R } = 10 ^{ - 2} $, and $ Q _\chi  = 0.1$. Also shown are the constraints from direct searches for $ Z 's $ and decays of $ \chi $ for the benchmark point chosen, as described in the text. The dashed grey contours show the level of fine-tuning necessary in our UV completion to avoid rapid $\chi \to \nu e^+ e^-$ and $ \chi \rightarrow \nu \nu \nu $ decays in tension with indirect detection bounds~\cite{Essig:2013goa,Gong:2008gi}.} 
\end{figure*}

We start by considering the regime $m_\chi \lesssim \text{ MeV}$. Since $E_R \propto 1/M$, lighter isotopes are particularly useful in probing such light candidates. Even still, reaching such light masses requires nuclear recoil thresholds lower than those of current experiments. We do not make detailed projections for any specific future experiment since they are diverse and would possibly involve absorption by collective modes rather than individual nuclei. Instead, we simply make projections for scattering off individual nuclei of Hydrogen or Lithium in future experiments with $E_{\text{th}}= 1 \text{ eV}$ and $M_T T = 100 \text{ kg yr}$ in Fig.~\ref{fig:NCsigmalight} (see \cite{Budnik:2017sbu,Szydagis:2018wjp} for proposals). The kink in the Lithium line is due to the two naturally occurring isotopes, \isotope[6]{Li} and \isotope[7]{Li}. Since $E_R \propto 1/M$ and we approximate the energy threshold with a step function, the kink occurs when the dark matter is too light to cause \isotope[7]{Li} to recoil with an energy above the threshold, but is still heavy enough to push \isotope[6]{Li} above the threshold. 

While the projected sensitivities only rely on the defining neutral current absorption operator in Eq.~\eqref{eq:NCops}, any UV completion will have other relevant constraints. In particular for our UV completion, setting $ m _{ Z '  } = 18~ {\rm GeV} $, $ s _{ \theta _R } = 10 ^{ - 1.5} $, and $ Q _\chi  = 0.1 $, the $\chi \to \nu \nu \nu$ decay is the most constraining indirect bound~\cite{Gong:2008gi}. Additionally, searches for such a $Z'$~\cite{Belyaev:2018pqr} and bounds on SM four-fermion interactions~\cite{Dror:2017nsg} are the most stringent direct constraints. Both direct and indirect bounds are shown as gray regions in the figure. We see that moderate, achievable energy thresholds and exposures of light isotopes can quickly probe viable, unexplored parameter space for $m_\chi \lesssim \text{ MeV}$.

Next, we consider heavier dark matter with $m_\chi \gtrsim \text{ MeV}$. This dark matter is sufficiently heavy to cause nuclear recoils above detector thresholds at existing experiments, as shown in Fig.~\ref{fig:NCsigma}. In fact, the recoil energies can be so large that they allow non-dark matter specialized experiments, such as Borexino~\cite{Agostini:2018fnx}, COHERENT~\cite{Akimov:2017ade,Scholberg:2018vwg} , and CUORE~\cite{Alduino:2017ehq}, to probe viable parameter space. We summarize the relevant details of each current experiment in Appendix \ref{sec:appExp}. Similar to our discussion of the kinked Lithium line, every kink in this figure corresponds to a particular isotope's recoil energy dropping below its corresponding experiment's recoil threshold. 

As discussed before, any particular UV completion of the neutral current absorption operator will have additional constraints. We set $ m _{ Z '  } = 18 ~{\rm GeV} $, $ s _{ \theta _R } = 10 ^{ - 2} $, and $ Q _\chi  = 0.1$ for our simple UV completion in the figure. Fine-tuning of $\epsilon$ in Eq.~\eqref{eq:NCdecaysee} is needed to avoid indirect detection bounds on $\chi \to \nu e^+ e^-$ decays~\cite{Essig:2013goa}, denoted with dashed grey contours labeled $\text{FT}_\epsilon$ in the figure. Additionally, fine-tuning against the IR contribution to the $\chi \to \nu \nu \nu$ decay in Eq.~\eqref{eq:NCdecaysThreeNu} is necessary to varying degrees~\cite{Gong:2008gi}, denoted with dashed grey contours labeled $\text{FT}_\nu$. These fine-tuning contours further motivate future iterations of current experiments with larger exposures and lower thresholds, such as Argo~\cite{Aalseth:2017fik}, DARWIN~\cite{Aalbers:2016jon}, PICO-500, and SuperCDMS SNOLAB~\cite{Agnese:2016cpb}. However, the requirement of fine-tuning to evade indirect detection bounds is highly model-dependent. For example, introducing flavor-dependent couplings might greatly reduce the need for any fine-tuning. Regardless, the projected bounds on $\sigma_{\text{NC}}$ are model-independent and encourage both searches for these neutral current fermionic dark matter absorption signals at current experiments and the study of UV completions of these operators which more naturally suppress decays bounded by indirect detection. 

\section{Charged Current: Induced $\beta$ Decays}
We now study the signals from the dimension-6 charged current operator generated by the UV model discussed in Sec.~\ref{subsec:CCmodel}:
\begin{equation} 
\label{eq:InducedBetaMinusExOp}
\frac{1}{\Lambda^2} \bigl[ \bar{p} \gamma _\mu \left( 1  + \lambda   \gamma _5  \right)  n  \bigr] \bigl[  \bar{e} \gamma ^\mu P _R \chi  \bigr] + \text{h.c.} \,,
\end{equation} 
where $  \lambda    \simeq  1.2694 \pm 0.0028$, with we identify $1/ \Lambda ^2 \equiv g _R ^2 /4M _{ W _R } ^2$.  For sufficiently massive dark matter, scattering on a nucleus can result in the conversion of a neutron/proton within the nucleus into a proton/neutron, accompanied by the emission of an energetic $e^{\mp}$ in the final state --- analogous to the familiar induced $\beta^{\mp}$ processes in neutrino physics.  This processes can lead to a variety of possible correlated signals depending on the target nucleus and dark matter mass. We now consider the scenario in which dark matter induces $\beta$ transitions in isotopes that are stable against $\beta$ decay in a vacuum. Signals arising from the decays of stable isotopes are particularly appealing as such nuclei exist in large abundances within the target material of current direct detection and neutrino experiments. We reserve study of unstable isotopes and the effect of dark matter transitions on the kinematic endpoint of their $\beta$ decay spectrum for Sec.~\ref{sec:CCEPshift}. 

Induced $\beta$ transitions will occur if the dark matter mass is above the kinematic threshold given by: 
\begin{align} 
\label{eq:betaminth}
m_\chi > m_{\rm th}^{\beta^{\mp}}  \equiv M_{A,Z\pm1}^\exc + m_e -  M_{A,Z} \, .
\end{align} 
Throughout this work, we take $M_{A,Z}$ to be the mass of the \emph{nucleus} of the isotope \isotope[A][Z]{X}. Note that if $M_{A,Z-1}<M_{A,Z} - m_e$, then the nucleus \isotope[A][Z]{X} can undergo electron capture. These isotopes are generally not long-lived, so the scenario of induced $\beta^+$ decay is most interesting for isotopes where electron capture is kinematically forbidden. Thus, for dark matter induced $\beta^+$ decay, we  limit ourselves to isotopes with $M_{A,Z-1}>M_{A,Z} - m_e$.  An additional complication occurs for $\beta^+$ decays in heavy isotopes where in general the number of neutrons is far greater than the number of protons, and Pauli Blocking effects would make $\beta^{+}$ transitions into the ground state or lowest lying excited states of the daughter nucleus disfavored. This motivates us to focus on signals of induced $\beta^+$ decays in experiments containing Hydrogen as a target. 

If the energy imparted upon the proton/neutron by the dark matter, \emph{i.e.} $ m _\chi $, is significantly less than the binding energy of the nucleus $\lesssim 10$ MeV, the proton/neutron will not have enough energy to escape and the dominant process will be from the outbound nucleon remaining bound to the nucleus, leading to two possible processes: 
\begin{align}
\label{eq:ntoporpton}
\nonumber
&\beta^- \text{:}  \quad \chi + n \rightarrow p + e^{-}  \,\, \Rightarrow \,\, \chi + \isotope[A][Z]{X} \rightarrow \isotope[A][Z+1]{X}^\exc + e^{-} \,, \\ 
&\beta^+ \text{:}  \quad \bar{\chi} + p \rightarrow n + e^{+}  \,\, \Rightarrow \,\, \bar{\chi} + \isotope[A][Z]{X} \rightarrow \isotope[A][Z-1]{X}^\exc + e^{+}\,.
\end{align}
Note that as a result of angular momentum conservation considerations, the daughter nucleus will generically be produced in an excited state. 
For dark matter masses greater than 10 MeV (where the incoming dark matter also begins to resolve the individual nucleons upon scattering), other signals are in general possible. In particular, with enough energy the incoming dark matter particle can break apart the nucleus and the ejected  proton/neutron will then hadronize and shower for energies above the QCD scale, leading to an array of possible new signals. The inclusive cross section in this case may be computed using standard techniques 
\cite{Formaggio:2013kya}. However, dark matter decays rates scale as a large power of $m_\chi$ (see Eq.~\eqref{eq:CCdecayrates} for the expressions for our charged current UV completion), and so will tend to be in conflict with astrophysical constraints at larger $ m _\chi $ for detectable cross-sections. We leave a detailed study of this regime to future work. 

The kinematics of the induced $ \beta $ process in the limit that $m_\chi \ll M_{A,Z}$ are straightforward to compute. The energy of the outgoing $e^\pm$ and daughter nucleus are: 
\begin{align}
\label{eq:CCkin}
E_{R }  \simeq \left\{ \begin{array}{lr}  m_\chi - \mCCth& {\rm (electron)} \\   \big( m_\chi -m_{\rm th}^{\beta} \big)^2/2 \mNprimePlus^{(*)} & {\rm (nucleus)} \end{array} \right.\,.
\end{align}
Note that the recoil energies parallel the neutral current case with the replacement $m_\chi \rightarrow m_\chi - m_{\rm th} ^{ \beta }$ for $m_\chi > m_e$. As can be seen from Eq.~\ref{eq:CCkin} the recoiling nucleus signal will be, as with the neutral current signal, velocity independent to leading order. 

We now compute the inclusive rate for dark matter induced $\beta$ decays and make projections for current experimental sensitivities. We defer discussion of specific signals to the end of this section but note that these charged current signals are typically well above experimental thresholds and have several correlated signals. Hence for most experiments of interest, the events are striking enough that they should easily pass experimental cuts. The signal rate for an experiment carrying a set of target isotopes parameterized by $ j $ is given by:
\begin{align}
\label{eq:CCrate}
R & = \frac{\rho_\chi}{2 m_\chi}  \sum_j N_{T, \, j}   \, n_j \langle \sigma v \rangle_j  \,,
\end{align}
where $\rho_\chi$ is the local dark matter density, and $N_{T, \, j}$ is the number of targets of a given isotope. Here we have assumed that any $e^{\pm}$ energy could potentially be detected, and integrate over all energies and angles of $e^{\pm}$ emission). An additional factor $n_j$ accounts for the total number of parton level targets for the  $\beta$ decay: 
\begin{align}
\label{eq:nj}
n_j  \simeq \left\{ \begin{array}{lr}  A_j - Z_j  & {\left(  \beta^{-} \right)} \\  \quad Z_j & {\left(  \beta^{+} \right) } \end{array} \right.\,.
\end{align}
We emphasize that the rate scales with target volume and experimental exposure. Therefore, computing the rate as in Eq.~\eqref{eq:CCrate} requires experimental input along with the computation of the scattering cross section for dark matter off nucleons. 

For $m_\chi $ below the binding energy of nucleons ($ m _\chi  \lesssim 10 \, \text{MeV}$), the absorption process cannot resolve the constituents of the nucleons and we consider only scattering off entire nuclei. In this regime, we may write the differential cross section (in the center of mass frame) as\footnote{\footnotesize{Note that to $\ord{v^0}$, the center of mass frame is approximately the lab frame, and we drop any indices to this affect.}} 
\begin{align}
\label{eq:CCdiffsigma}
 \frac{d \, \sigma}{d \, \Omega } 
%= \frac{1}{32 \pi M^2} \frac{|\vec{p}_3|}{\beta_\chi E_\chi} \frac{1}{4} \sum |\mathcal{M}|^2
= \frac{1}{64 \pi^2 E_{\text{cm}}^2 } \ \frac{|\vec{p}_e|}{|\vec{p}_\chi|}  \sum_{\text{transitions}}   \overline{\magn{\MM_N}^2}  \, ,
\end{align}
where $\Omega$ is the solid angle $\theta$ is the angle between the incoming dark matter and the emitted electron/positron, $\mathcal{M}_N$ the amplitude is for scattering off of nucleons, $ \vec{ p } _{e} $ ($ \vec{ p} _\chi $) is the momentum of the electron (dark matter), and here we sum over possible nuclear spin states which manifests as a sum over the allowed transitions. Note that Eq.~\eqref{eq:CCdiffsigma} holds for both induced $\beta^{-}$ and $\beta^{+}$ decays. However, the nuclear rate will depend on which of the two processes is under-consideration, and we discuss this further in what follows.  

We emphasize that any large exposure experiments (both neutrino and designated dark matter direct detection) can be re-purposed to search for induced beta decays from fermionic absorption provided the kinematic threshold, Eq.~\eqref{eq:betaminth},  is low enough. A list of select stable (or meta-stable) isotopes in which light dark matter could induce $\beta$ transitions is summarized in Table~\ref{tab:isotopes}. The isotopes are grouped by their stability against $\beta^-$ or $\beta^+$ decay, and for each possible transition we quote the value of the threshold given by Eq.~\eqref{eq:betaminth} for the $\Delta I = 0$  and $\Delta  I = \pm 1$ transition with the lowest threshold --- (though other transitions are also generally accessible --- see Fig.~\ref{fig:threholdvsA} for a summary of various transitions corresponding to the experimental targets we consider here). 

\begin{table}
\center
\bgroup
\def\arraystretch{1.4}
\begin{tabular}{lc ll}
\toprule[1.5pt]
&   Process    &  Isotope (Threshold $\Delta I = 0$, $\Delta I = \pm 1$ )    \\ 
\midrule[1pt]  
$ \beta^-$: 
& $  \prescript{A}{6}{\text{C}}\rightarrow \prescript{A}{7}{\text{N}}    $ & $  \prescript{12}{6}{\text{C}} ( 18.3~{\rm MeV} ,\, 17.3~\rm{MeV} )$, $  \prescript{13}{6}{\text{C}} (2.22 ~{\rm MeV} ,\, 5.7~\rm{MeV} )$   \\ 
 & $  \prescript{A}{8}{\text{O}}\rightarrow \prescript{A}{9}{\text{F}}    $ & $\prescript{16}{8}{\text{O}} (16.4~{\rm MeV} ,\,  16.4~\rm{MeV})$, $\prescript{17}{8}{\text{O}} (2.76 ~{\rm MeV} ,\,  3.75~\rm{MeV})$,  \\ 
 &&  $\prescript{18}{8}{\text{O}} (2.70 ~{\rm MeV} ,\,  1.65~\rm{MeV})$   \\ 
  & $  \prescript{A}{52}{\text{Te}}\rightarrow \prescript{A}{53}{\text{I}}  $ & $  \prescript{130}{52}{\text{Te}} (1.42~{\rm MeV},\, 6.62~{\rm MeV})$, $  \prescript{128}{52}{\text{Te}} (2.25~{\rm MeV},\, 1.25~{\rm MeV})$, \\ & &    $  \prescript{126}{52}{\text{Te}} (3.1~{\rm MeV},\, 1.41~{\rm MeV})$,  $ \prescript{125}{52}{\text{Te}} (430~{\rm keV},\, 374~{\rm keV})$    \\   
 & $  \prescript{A}{54}{\text{Xe}}\rightarrow \prescript{A}{55}{\text{Cs}}    $ & $\prescript{129}{54}{\text{Xe}} (1.19~{\rm MeV},\, 1.33~{\rm MeV})$, $\prescript{131}{54}{\text{Xe}} (570~{\rm keV},\, 355~{\rm keV})$, \\ & & $\prescript{134}{54}{\text{Xe}} (2.23~{\rm MeV},\, 490~{\rm keV})$, $\prescript{136}{54}{\text{Xe}} (1.09~{\rm MeV},\, 1.06~{\rm MeV})$    \\ 
\midrule[1pt]  
$\beta^+$:  
  & $ \prescript{1}{1}{\text{H}}   \rightarrow  n  $ &   $ \prescript{1}{1}{\text{H}} (1.8~{\rm MeV}) $  \\   
\bottomrule[1.5pt]
\end{tabular}
\egroup
\caption{\footnotesize{Here we summarize notable isotopes that could undergo induced $\beta$ transitions, with a focus on the most abundant isotopes of materials interesting for the experiments we consider here (see Appendix~\ref{sec:appExp}). The right hand column displaces the threshold Eq.~\eqref{eq:betaminth} for the transition with the smallest splitting between the ground state of the parent nucleus and the excited state of the daughter allowed by the selection rule in question $(\Delta I = 0, \Delta I = \pm 1)$. Note that for Gamow Teller $\Delta I = \pm 1$ ground state to ground state transitions are possible leading to the sub-keV thresholds. For induced $\beta^{-}$ transitions many processes exist with low threshold (see Fig.~\ref{fig:threholdvsA} for additional possible higher threshold transitions), however for $\beta^{+}$ transitions Pauli blocking effects in heavy isotopes will generally disfavor transitions into the lowest lying excited states of the daughter nucleus. This leads us, in the present work, to consider only induced $\beta^{+}$ processes involving target materials with Hydrogen.}}
\label{tab:isotopes}
\end{table}

The plethora of different isotopes that undergo induced $\beta^{-}$ transitions leads to a wide range of signals at various experiments utilizing different target materials.  
 For a list of experiments and corresponding target materials and exposures, see appendix~\ref{sec:appExp}.  From Table~\ref{tab:isotopes} we see that the lowest possible dark matter mass that the induced beta decay signal can probe is about $m_\chi \sim 355$ keV from absorption by a $ \prescript{131}{54}{\text{Xe}}$ nucleus at a Xenon based experiment. Additionally note that $\prescript{125}{52}{\text{Te}}$ has a particularly low threshold of $374$ keV, making CUORE, which utilizes $\text{Te} \text{O}_2$ crystals and was designed to search for neutrino-less double $\beta$ decay, particularly suited to probe sub-MeV dark matter masses. For the charged current process, we focus entirely on current experiments. Interestingly, there is one transition in the Standard Model with an anomalously small threshold of $2.5  ~{\rm keV}$ that can employ $\Delta I = \pm 1$ transitions, $\chi + \prescript{163}{66}{\text{Dy}}   \rightarrow \prescript{163}{67}{\text{Ho}} + e^{-}$. This was considered in ~\cite{Lasserre:2016eot} as a way to probe sterile neutrinos. Unfortunately, the tremendous expense of building large volume experiments filled with Dysprosium make it a challenging direction to observe fermionic absorption.
 
\subsection{Induced $\beta^{-}$ Decays}
We now focus specifically on signals from induced $\beta^{-}$ decays
\begin{align} 
\chi (m_\chi \, \vec{v}) + \isotope[A][Z]{X} (\vec{0}) \to e^{-}(\vec{p}_e) +  \isotope[A][Z+1]{X} (\vec{q}) \,,
\end{align} 
and compute projected experimental limits by computing the expected rate Eq.~\eqref{eq:CCrate} at specific experiments given current exposures.  The nucleon level amplitude, required to compute the differential scattering cross section in Eq.~\eqref{eq:CCdiffsigma}, may then be written as follows: 
\begin{align}
\label{eq:CCamp}
\mathcal{M}_{N}= \sqrt{\mathcal{F}(Z+1, E_e)}   \mathcal{M}  \,.
\end{align}
Here $\mathcal{M}$ is the parton level scattering amplitude for $\beta^{-}$ transitions $\chi + n \rightarrow p + e^{-}$  generated by the operator in \eqref{eq:InducedBetaMinusExOp}, with the $p$/$n$ momentum normalized to nuclei mass \emph{i.e.} $p^\mu p_\mu = M_{A_j, \, Z_j}^2$\footnote{\footnotesize{Note that the vector and axial vector form factors are implicitly defined in \eqref{eq:InducedBetaMinusExOp}} for low momentum transfer $q^2 = (k_\chi - k_e)^2$, and agree with results from the neutrino literature namely $f_V(q^2 \sim 0 ) = 1$ and  $f_A(q^2 \sim 0) = -1.2694$ \cite{Formaggio:2013kya}. Note that while the technology exists to compute form factors for scattering between different nuclei generated by general operators, we are unaware of such a computation carried out in the literature.}. The factor $\mathcal{F}(Z + 1,E)$ in Eq.~\eqref{eq:CCamp} is the usual Fermi function accounting for Coloumb interactions between nucleons \cite{Formaggio:2013kya} for $\beta$ transitions, and is given by:
\begin{align}
& \mathcal{F}(Z,E_e) = 2 (1+S) \frac{| \Gamma (S + i \eta) |^2 }{ \Gamma ( 1 + S)^2} \bigl(2  r_N |\vec{p}_e | \bigr)^{2 S - 2} e^{\pi \eta} \, , 
\end{align}
where $\eta = \alpha Z E_e / | \vec{p}_e | = \alpha Z E_e /\sqrt{E^2_e - m_e^2} $ and $S = \sqrt{1 - \alpha^2 Z^2}$. Here the nuclear radius is given by $r_N = 1.2 ~{\rm fm}\, A^{1/3}$. For large $ E _e \gg m _e $, the Fermi function asymptotes to a larger for isotopes with larger $Z$ --- therefore experiments utilizing light target nuclei (such Super-Kamiokande and Borexino) are particularly sensitive at larger dark matter masses (as they result in higher energy electrons). 

\begin{figure} 
\centering 
\hspace{-0.3cm}
\includegraphics[width =.72\textwidth]{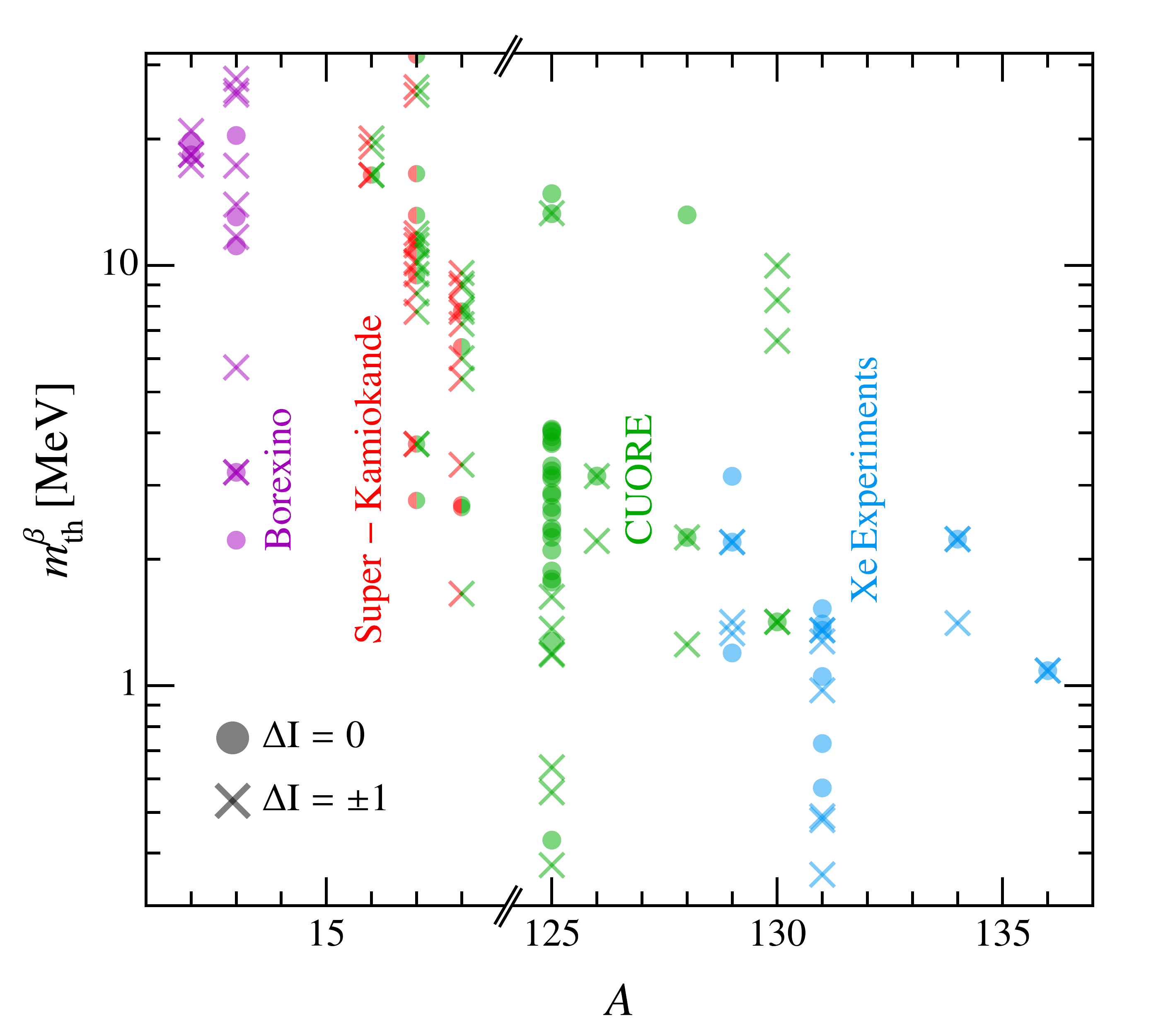}
\caption{A comprehensive summary of every induced $\beta^-$ decay we consider, plotted in the $A$ versus $m_{\text{th}}^\beta$. Circles represent nuclear transitions in which the nuclear spin does not change, while crosses represent those which change by 1. Note that Super-Kamiokande and CUORE both contain oxygen, hence their overlap from $A=16-18$. }
\label{fig:threholdvsA}
\end{figure}

As discussed above, Eq.~\eqref{eq:CCdiffsigma} includes a sum over all possible nuclear spin states. The Lorentz structure of a given charged current operator dictates what kind of angular momentum selection rules are in effect, and therefore which nuclear transitions are allowed. For the model considered here the operator in Eq.~\eqref{eq:CCdiffsigma} contains both vector and axial vector couplings. In the case of a pure vector operator and light dark matter, Fermi transitions will dominate (other transitions will be suppressed by factors of $e^{-m _\chi r _N }$). Fermi transitions are transitions in which the spin of the dark matter and the electron are parallel so that the spin angular momentum of the initial and final nucleus is unchanged $\Delta I_F = 0$.  Meanwhile, for axial and axial-vector couplings, Gamow-Teller transitions become possible and contribute to the sum. In these transitions the dark matter and electron have anti-parallel spins so that the nucleus spin must change to conserve angular momentum $\Delta I_{GT} = 1$. Both Fermi and Gamow-Teller transitions preserve parity ($ \pi $). Note that here we have focused entirely on the low energy limit. For $ m _\chi \gg r _N ^{-1} $ all angular momentum transitions are accessible. A summary of all possible Fermi and Gamow-Teller transitions for the experimental target materials considered here is presented in Fig.~\ref{fig:threholdvsA}, where we also show the kinematic threshold for $\beta^-$ decays to occur given each excited state mass $M_{Z_j+1, A_j}^{(*)}$  as dictated by Eq.~\eqref{eq:betaminth} (the smallest thresholds possible for Fermi and Gamow-Tellar transitions is also quoted in Table.~\ref{tab:isotopes})

%\newpage
The parton spin averaged nucleon level matrix element arising from the charged current operator Eq.~\eqref{eq:InducedBetaMinusExOp} for induced $\beta^{-}$ decay is given by: 
\begin{align}
\label{eq:partonmatrixsquared}
\overline{\magn{\MM}^2} = &  \frac{4 m _n m _\chi}{ \Lambda^4} \bigg[ E_e \left( 2 m_{n}  -m_{p} + 2m_\chi-E_e \right)  -m_e^2 \notag \\ 
& \qquad +2 \lambda \left( E_e^2-m_e^2 \right)    +\lambda^2 \left( E_e \left(  2m_{n} +m_{p} +2m_\chi-E_e \right)  -m_e^2 \right) \bigg] \,,
\end{align}
 Note that Eq.~\eqref{eq:partonmatrixsquared} contains both vector ($\propto \lambda^0$), axial vector ($\propto \lambda^2$), and interference terms ($\propto \lambda$). All terms contribute to $\Delta I = 0$ and $\Delta \pi = 0$ transitions, while the axial term additionally allows for Gamow-Teller transitions $\Delta I = \pm 1$ and $\Delta \pi = 0$. Therefore, the daughter nucleus is often necessarily be formed in an excited state --- one that obeyed the Fermi or Gamow-Teller spin angular momentum selection rule. 

With Eq.~\eqref{eq:partonmatrixsquared} we can compute thermally averaged cross section from Eqs.~\eqref{eq:CCdiffsigma} - \eqref{eq:CCamp} for induced $\beta^{-}$ decays, which we find to be: 
\begin{align}
\label{eq:CCsigma}
\langle \sigma v \rangle_j =  \frac{|\vec{p}_e|_j}{ 16 \pi m_\chi  \,\mNj^2} \overline{  |\mathcal{M} _{N_j} |^2} \,,
\end{align}
where as the rate is independent of solid angle to leading order we able to trivially carry out the angular integral. Here, $ |\vec{p}_e |^2_j = (\mCCthj - m_\chi)(\mCCthj - m_\chi - 2 m_e)$  is the electron's outgoing 3-momentum in the center of mass frame (which is approximately the lab frame), in the limit that $m_e, \, m_\chi, \mCCthj \ll \mNj$. The $j$ index refers to a specific transition (Fermi or Gamow-Teller) of a given isotope of an experimental target material. 
The total event rate for induced $\beta^{-}$ may now be computed by plugging in Eq.~\eqref{eq:CCsigma} into Eq.~\eqref{eq:CCrate} and summing over the contributions from each isotope and each possible transition a given isotope could undergo under the angular momentum selection rules:
\begin{align}
\label{eq:CCrateexplicit}
R & = \frac{\rho_\chi}{2 m_\chi}   \sum_j N _{ T, j}   n_j \frac{|\vec{p}_e|_j}{ 16 \pi m_\chi  \,\mNj^2} \mathcal{F}(Z+1, E_e)  \overline{\magn{\MM}^2}  \,.
\end{align}
In contrast to the neutral current signal rate there are no implicit experimental cuts imposed on the induced $\beta^{-}$ event rate at this level \emph{i.e.} we assume all events are above the experimental energy threshold (with the notable exception being Super-Kamiokande which has a higher energy detection threshold of 3.5 MeV~\cite{Wan:2019xnl}). 

We now comment on the possible signals due to a charged current event. From the kinematics in Eq.~\eqref{eq:CCkin} we see that there is significant velocity-independent nuclear recoil energy, analogous to the situation in neutral current processes. One can then look for correlated peaks between different isotopes in the target material as discussed in Sec.~\ref{sec:neutral}. Even more striking, charged current processes result in an emitted energetic $e^{\pm}$ which could be observed at experiments. The emitted energetic electron can shower in the detector, and one may then search for this electron in parallel with the nuclear recoil. Additionally, the daughter nucleus will typically be produced in an excited state (as determined by angular momentum selection rules) which will then decay (typically with a known lifetime). This secondary decay will emit a photon which may be searched for at experiments sensitive to photon emission. Furthermore, the produced nucleus itself may be unstable and decay on time scales of interest to the experiment. In summary the possible signals are summarized as follows:
\begin{itemize}
\item{Emitted high energy electron}
\item{Recoiling nucleus at recoil energies peaked around a single bin (analogous to the neutral current case discussed above).}
\item{Photon from nucleus being produced in an excited state}
\item{Decay of unstable nucleus}
\end{itemize}
The plethora of correlated signals makes it possible to trigger on several different signals. The optimal search strategy will depend on the specific target being considered, experimental capabilities, and the dark matter mass. 

As an example consider $\beta^{-}$ induced decays within a Xenon based detector. Stable (or meta-stable) isotopes can undergo the charged current absorption process:
\begin{equation}
 \chi + \prescript{A}{54}{\text{Xe}} \rightarrow e^{-}  +  \prescript{A}{55}{\text{Cs}}^{(*)}  \,,
\end{equation}
where $A =  \{ 126 \, (28.4 \%),$ $131 \, (21.2 \%),$ $134 \, (10.4 \%),$ $136 \,(8.8 \%) \} $ are the dominant isotopes given in their natural abundance (note that neutrino-less double beta decay experiments use enriched $\prescript{136}{54}{\text{Xe}}$). Depending on the Xenon isotope and the transition that takes place, the resultant Cesium nucleus may or may not be in an excited state. As an explicit example, consider $\prescript{131}{54}{\text{Xe}}$. The ground state of this isotope is a $I^{P} = \frac{3}{2}^{+}$ state while that of $\prescript{131}{55}{\text{Cs}}$ is in a $\frac{5}{2}^{+}$ state. The Fermi $\Delta I_F = 0$ transition $\frac{3}{2}^{+} \rightarrow \frac{3}{2}^{+}$ produces an excited state of Cesium 60 keV above the Xenon ground state. Adding the electron mass results in a threshold of $m_{\rm th}^{\beta} = 571$ keV for the process to occur. 

\begin{figure} 
\centering \hspace{-1.6cm}
\includegraphics[width =.78\textwidth]{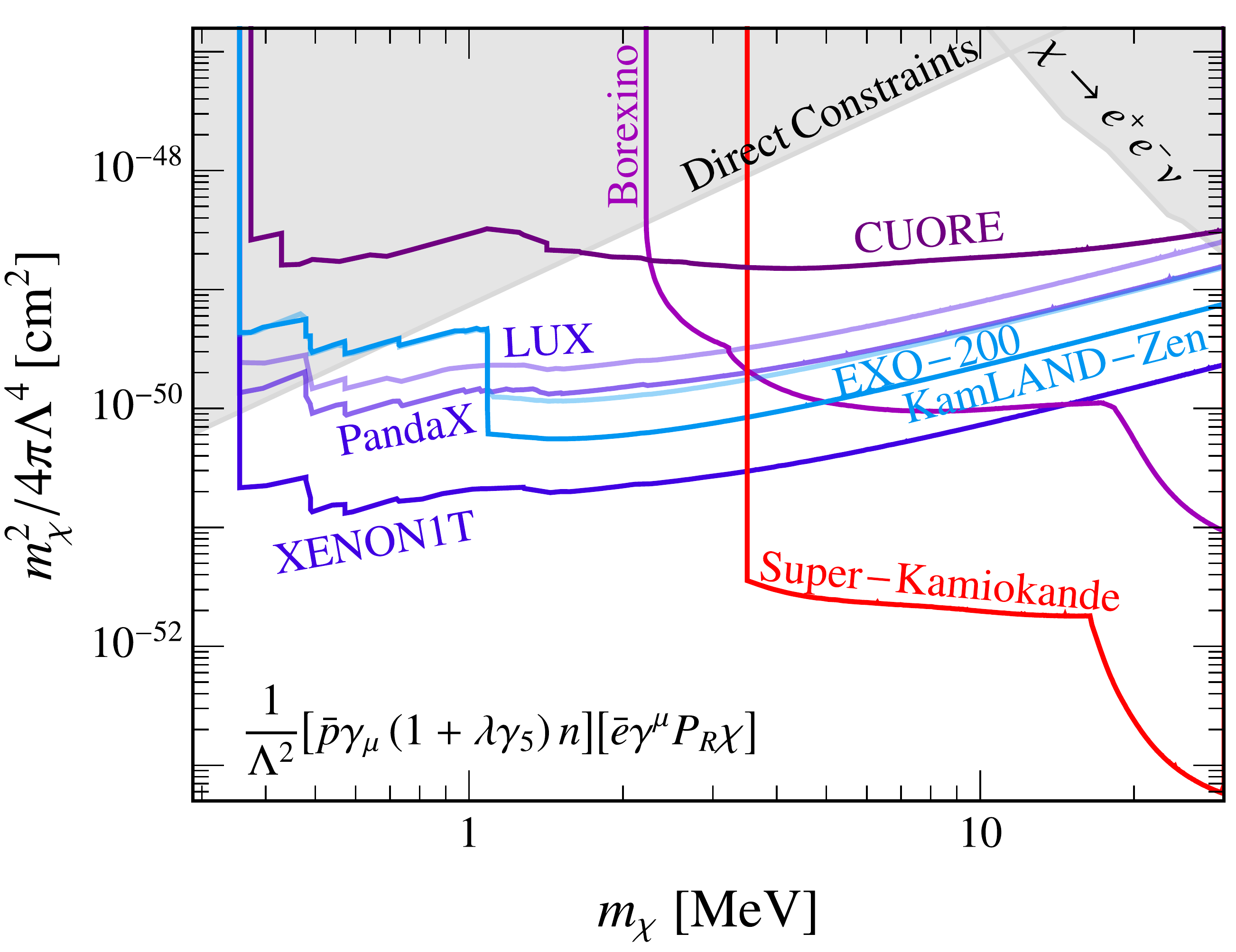}
\caption{The projected constraints from a dedicated search for induced $\beta^-$ signals at XENON1T, LUX, Panda-XII, EXO, and KamLAND-Zen for Xenon conversion. Super-Kamiokande and Borexino for absorption  by Hydrogen, and by $\text{Te} \text{O}_2$ crystals in CUORE. The various experiments, their exposures and physics goals are summarized in Appendix~\ref{sec:appExp}.
 }
\label{fig:betaminusProjections}
\end{figure}

For an axial vector coupling the Gamow-Teller, $\Delta I_{GT} = 0, \,  \pm1$ transitions are also possible and one must take into account the contribution of all such possible transitions in the amplitude.  In particular, the $\frac{3}{2}^{+} \rightarrow \frac{5}{2}^{+}$ transition from the ground state of $\prescript{131}{54}{\text{Xe}}$ to the ground state of $\prescript{131}{54}{\text{Cs}}$ contributes with threshold 360 keV, additionally transitions to the third excited state of $\prescript{131}{54}{\text{Cs}}$ also occur at relatively low threshold 490 keV.  For setting limits, we sum over all relevant contributions from the various allowed transitions. Fig.~\ref{fig:threholdvsA} shows the values of the various $\Delta I = 0$ and Gamow-Teller only $\Delta I = \pm 1$ transitions for a given isotope atomic number that correspond to relatively low thresholds and significant abundances.

To project the sensitivity of current experiments to charged current signal we require at least 10 events with the results shown in Fig.~\ref{fig:betaminusProjections} for various experiments summarized in Appendix~\ref{sec:appExp}, where note that again (with the exception of Super-Kamiokande) these projections assume no experimental cuts which is motivated due to the large number of correlated possible signals; nuclear recoil, energetic $e^{-}$ (searches for high energy electrons could in principle even be done using an existing analysis, for instance the S2 XENON1T data set \cite{Aprile:2016wwo}, but we leave a detailed analysis to future work), emitted $\gamma$, and decay of an unstable daughter nucleus. For some isotopes, the excited states have not yet been fully mapped out. In general, however for heavier elements an excited state with matching angular momentum for each transition should exist within $ \sim $ MeV of the ground state. For practical purposes when the data on the excited state corresponding to the $\Delta I = 0$ transition is not available we take the splitting to be 1 MeV. 

As with the neutral current case, limits are sensitive to the different isotopes in a given experiment, where $m_{\rm th}^\beta$ for all the possible transitions in the experiments considered here are summarized in Fig.~\ref{fig:threholdvsA}. In particular, the discontinuities in the limits of Fig.~\ref{fig:betaminusProjections} occur at $m_\chi \sim m_{\rm th}^\beta$, below which transitions become inaccessible for a given isotope. Projected scale linearly with exposure of a given experiment as is expected from \eqref{eq:CCrate}. In particular, consider the projections for searches for induced $\beta^-$ signals at XENON1T, LUX, Panda-XII, EXO, and KamLAND-Zen for Xenon conversion. Note that KamLAND-Zen and EXO are dedicated neutrino-less double beta decay experiments which utilize enriched  $\prescript{136}{54}{\text{Xe}}$ target material, while Xenon based dark matter detectors contain Xenon isotopes in their natural abundance within. As expected the projected limits scale with exposure with XENON1T being more sensitive than LUX or Panda-XII. At high energies EXO and KamLAND-Zen do better than LUX and Panda-XII due to their larger exposures. However once energies fall below the threshold for induced beta decays by absorption off $\prescript{136}{54}{\text{Xe}}$ at about an 1 MeV, absorption occurs through the significantly sub-dominate isotopes, weakening the projected limits. CUORE, an experiment employing $\text{Te} \text{O}_2$ crystals looking for neutrino-less double beta decay, is also shown and could search for absorption of dark matter off both the Tellurium and Oxygen nuclei. 
Note that Super-Kamiokande and Borexino do particularly well simply due to their enormous Hydrogen detector volume and in particular enjoy an enhancement at large energies due to the Fermi function relative to detectors with larger $Z$ target materials. The capabilities of Super-Kamiokande become more pronounced for $m_\chi \gtrsim  16 \, \text{MeV}$ when it becomes kinematically possible to induce $\beta^{-}$ decays off \isotope[16][8]{O} (which comprises about $99 \%$ of the oxygen in its natural abundance).  

In Sec.~\ref{subsec:CCmodel} we discussed constraints from colliders as well as  indirect detection constraints due to dark matter stability for the case of the charged current UV model at hand. Regions excluded due to these constraints are shaded out, and as expected stability considerations favor dark matter masses lighter than about 10 MeV and collider constraints favor mediator scales heavier than about a TeV. 

\subsection{Induced $\beta^{+}$ Decay}
We now briefly consider signals from induced $\beta^{+}$ decays. In this process, an incoming dark matter converts a proton into a neutron and a positron. For heavy isotopes whose nuclei contain a significantly larger fraction of neutrons over protons, Pauli blocking of the outgoing neutron disfavors the production of a neutron in the lowest-lying energy states, thereby resulting in larger thresholds. Additionally, while such transitions have been studied in the context of supernova neutrinos, a detailed analysis of the favored transitions is not available in the literature and we leave this to future work. For the present, we simply focus on the case of induced $\beta^+$ transitions in experiments employing a target material consisting of Hydrogen, and consider the process;
\begin{equation}
 \chi + \prescript{1}{1}{\text{H}} \rightarrow n + e^{+}  \,,
\end{equation}
with threshold of $1.8$ MeV. 

\begin{figure} 
\centering \hspace{-0.8cm}
\includegraphics[width =.78\textwidth]{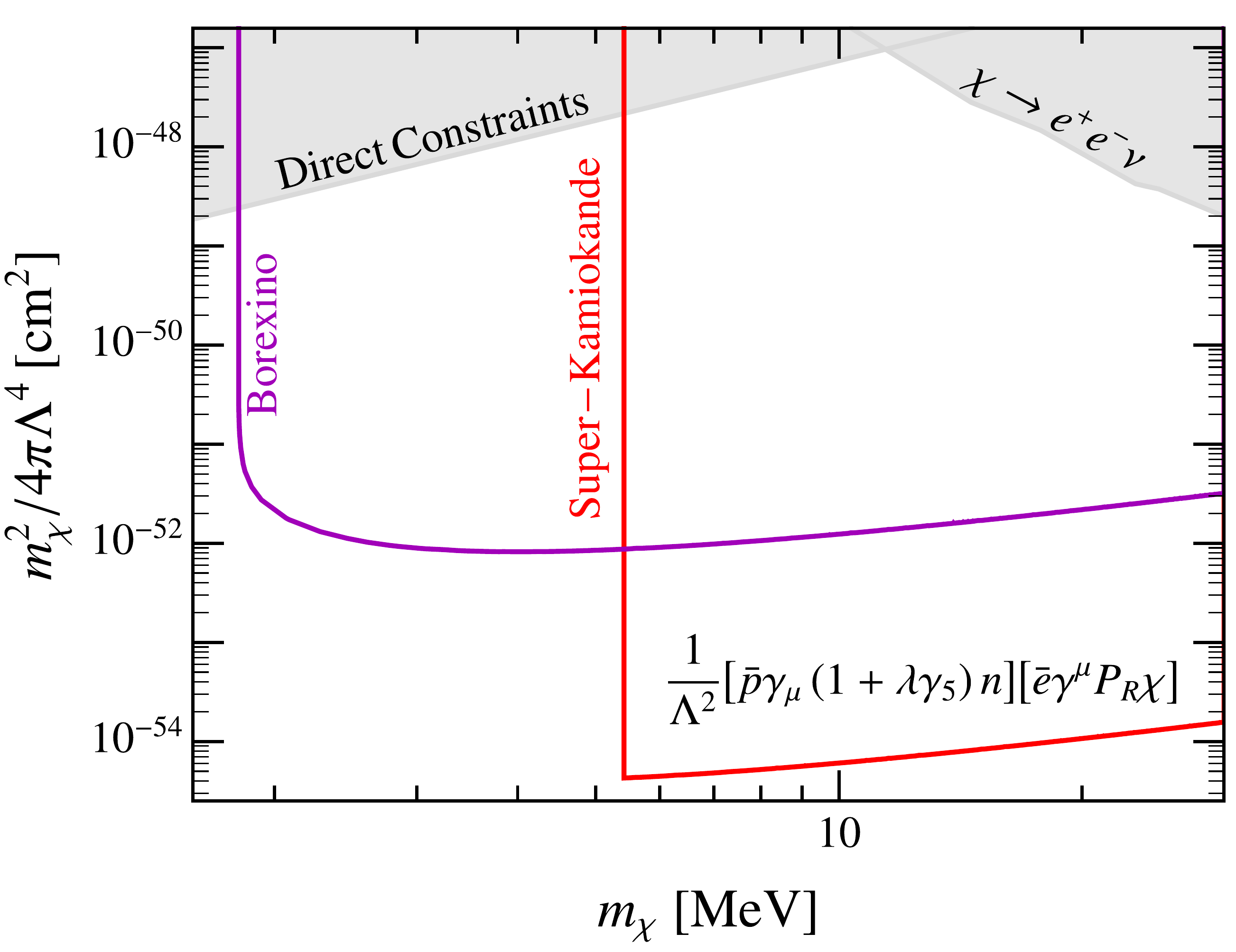}
\caption{The projected constraints from a dedicated search for induced $\beta^+$ signals from Hydrogen at Super K and Borexino.}
\label{fig:betaplusProjections}
\end{figure}

The kinematics, matrix element and rate can be calculated from the discussion above.  In Fig.~\ref{fig:betaminusProjections} we show projected limits for induced $\beta^+$ decay in the liquid $\text{H}_2 \text{O}$ of Super-Kamiokande, and the $\text{C}_6 \text{H}_3 (\text{C}\text{H}_3)_3$ target material of Borexino. In this way, two large volume experiments can now probe low dark matter masses --- down to the kinematic threshold of $ 1.9 \text{ MeV}$ in Borexino (which has a detection threshold of $ 70 ~{\rm keV} $), and the $5.3$ MeV in Super-Kamiokande (which has a detection threshold of $ 3.5 ~{\rm MeV} $). Projected limits for these to experiments are shown in Fig.~\ref{fig:betaplusProjections}, once again, to set the limit, we assume that a given experiment can resolve and measure the energy of the produced $e^{\pm}$. 
Relative to $\beta^{-}$, $ \beta ^+ $ processes tend to have larger thresholds leading to complementarity between the two types of searches. Lastly we note that asymmetric dark matter models may result in only $ \beta ^- $ or $ \beta ^+ $, further motivating carrying out both types of searches.

\section{Charged Current: $\beta$ Endpoint Shifts} 
\label{sec:CCEPshift}
Having already considered $\beta$ transitions induced by charged current operators, we now consider the possible signals in isotopes which $\beta$ decay without the presence of dark matter. In these isotopes, light dark matter absorption with an unstable parent nucleus causes a shift in the kinematic endpoint of the $\beta$ spectrum. Since the decay is allowed in vacuum, this is a threshold-less process and can occur for arbitrarily light dark matter. In particular, kinematic endpoint shifts can probe $m_\chi \lesssim \text{ MeV}$, where induced $\beta$ transitions are kinematically forbidden by Eq. \eqref{eq:betaminth}. Isotopes which $\beta^+$ decay have smaller scattering rates than those which $\beta^-$ decay due to the Fermi function, while those which electron-capture decay still have kinematic thresholds (albeit smaller ones). Thus, we ignore targets which naturally $\beta^+$ or electron-capture decay and just focus on those which $\beta^-$ decay\footnote{One could also consider re-purposing sterile-neutrino search experiments which used $\beta^+$ \cite{Trinczek:2003zz} and electron-capture~\cite{Hindi:1998ym} decaying  isotopes. Unfortunately, the smallest mixing angle they constrain is $4\times 10^{-3}$ \cite{Trinczek:2003zz} which corresponds to an already-constrained $\Lambda$ in the charged current operator in Eq.~\eqref{eq:InducedBetaMinusExOp} from direct searches.}. 

Detecting rare $\beta^-$ spectra endpoint shifts is difficult due to the lack of experiments with large exposures of unstable targets. There are only a few well motivated physics goals which employ $\beta^-$ decaying isotopes: measuring the SM neutrino masses~\cite{Wolf:2008hf,Nucciotti:2010tx,Esfahani:2017dmu}, detecting the Cosmic Neutrino Background (C$\nu$B)~\cite{Betti:2019ouf}, and producing light sterile neutrinos~\cite{deVega:2011xh,Smith:2016vku}. In this section we consider the possibility of using one of these existing or proposed experiments to look for fermionic absorption. Needless to say, the experimental requirements for detecting dark matter in this way are often quite relaxed relative to those needed for other experimental physics goals. For example, C$\nu$B detection experiments must resolve $\beta^-$ energies at the possible scale of SM neutrinos' masses, $m_\nu \sim 0.1 \text{ eV}$, thus requiring a target with a low $Q$ value: tritium. By contrast, the fermionic dark matter we consider is heavier: $m_\chi \gtrsim 190 \text{ eV}$~\cite{DiPaolo:2017geq,Savchenko:2019qnn}, so this target selection criterion is irrelevant. This motivates us to consider different $\beta^-$ decaying isotopes more generally than the present proposals. 

Toward this end, we categorize every isotope which dominantly $\beta^-$ decays and has a half-life between $10^8$ and $10^{13}$ seconds in Table \ref{tab:betaisos}\footnote{We do not consider shorter half-lives since tritium's half-life is roughly $4 \times 10^8$ seconds and would therefore be a better target than any shorter lived isotope. Isotopes with longer half-lives than $10^{13}$ seconds decay via even higher order forbidden transitions.}. In addition to dominantly decaying via Fermi or Gamow-Teller transitions, some of these isotopes predominantly undergo first forbidden transitions ($\Delta \pi =1$, $\Delta I =0,\pm 1, \pm 2$) or second forbidden transitions ($\Delta \pi =0$, $\Delta I =\pm 2, \pm 3$), where $\Delta \pi$ is the parity change and $\Delta I$ the spin change.
\begin{table*}
\center
\bgroup
\def\arraystretch{1.5}
\setlength{\tabcolsep}{2pt}
\begin{tabular}{l|cccccccccccc}
\toprule[1.5pt]
Fermi &  \isotope[3]{H}   \\ 
\midrule[1pt] 
Gamow-Teller & \isotope[14]{C} & \isotope[32]{Si} & \isotope[60]{Co} & \isotope[63]{Ni} & \isotope[154]{Eu} & \isotope[228]{Ra} \\
\midrule[1pt] 
1st forbidden & \isotope[39]{Ar} & \isotope[42]{Ar} & \isotope[79]{Se} & \isotope[85]{Kr} & \isotope[90]{Sr} &  \isotope[137]{Cs} & \isotope[151]{Sm} & \isotope[194]{Os} & \isotope[204]{Tl} & \isotope[210]{Pb} & \isotope[227]{Ac} & \isotope[241]{Pu} \\
\midrule[1pt]
2nd forbidden & \isotope[36]{Cl} & \isotope[94]{Nb} & \isotope[99]{Tc} & \isotope[126]{Sn} \\
\bottomrule[1.5pt]
\end{tabular}
\egroup
\caption{Every isotope which dominantly $\beta^-$ decays and has a half-life between $10^8$ and $10^{13}$ seconds, categorized by their main transition~\cite{NuDat}. Only Fermi and Gamow-Teller transitions are not momentum-suppressed for our charged current operators and therefore of interest to us.
}
\label{tab:betaisos}
\end{table*}
The Fermi and Gamow-Teller transition isotopes in Table~\ref{tab:betaisos} are of particular interest to us since they are the most long lived $\beta^-$-decaying isotopes for which the dark matter capture rates would not be momentum suppressed. We systematically checked the most recent $\beta^-$-decay experiments to verify that none had significant exposures of these interesting targets\footnote{References for each isotope are \isotope[14]{C}~\cite{Kuzminov2000}, \isotope[32]{Si}~\cite{AUDI2003337}, \isotope[60]{Co}~\cite{Hansen1968}, \isotope[63]{Ni}~\cite{OLSSON199277}, \isotope[154]{Eu}~\cite{NG1968433}, and \isotope[228]{Ra}~\cite{Wang_2012}. }. This is expected since the exposures required to determine $\beta^-$ spectra and half-lives are generally small. 

The largest experimental proposals with targets which can undergo a Fermi or Gamow-Teller transition are made of tritium. The proposal with the largest tritium exposure is PTOLEMY which hopes to be the first experiment to measure the C$\nu$B~\cite{Betti:2019ouf}\footnote{While KATRIN~\cite{Wolf:2008hf} and Project 8~\cite{Esfahani:2017dmu} also use tritium, they need far less than PTOLEMY's proposed amount for their measurements of the electron antineutrino's mass and will not probe charged current operator parameter space which is not already ruled out by LHC constraints.}. There has also been recent interest in using a comparable exposure of tritium in an experiment to measure coherent neutrino-atom scattering \cite{Cadeddu:2019qmv}. Regardless, we will do a proposal-independent analysis below when projecting sensitivities to the charged current signal which only depends on the exposure of tritium.

Tritium $\beta^-$ decays to Helium via
\begin{align} 
\h3 \to \he3 + e^- + \bar{\nu}_e,
\end{align} 
where $\h3$ and $\he3$ refer to the nuclei (and not the atoms) of those isotopes. The Q-value for this decay is
\begin{align} 
Q = m_{\h3}-m_{\he3}-m_e -m_\nu \simeq  18.6 \text{ keV},
\end{align} 
where $m_{\h3}\simeq 2808.921 \text{ MeV}$ and $m_{\he3}\simeq 2808.391 \text{ MeV}$ are the nuclear masses. The charged current operators in Eq.~\eqref{eq:InducedBetaMinusExOp} allow tritium to capture incoming dark matter via
\begin{align} 
\chi + \h3 \to  \he3 + e^-.
\end{align} 
Since the tritium nuclear transition is from $1/2^+ \to 1/2^+$, both the vector and axial-vector operators contribute to the fermionic absorption rate. 

Using Eqs.~\eqref{eq:CCrate} and \eqref{eq:CCsigma} from above, we find the rate for tritium to absorb fermionic dark matter is
\begin{align} 
R & = \frac{\rho_\chi}{ m_\chi} N_{\isotope[3]{H}}  \frac{\magn{\vec{p}_e}}{4\pi m_{\isotope[3]{H}} \Lambda^4} F \left( Z+1,E_e \right)  \bigg[ E_e \left( 2 m_{\isotope[3]{H}}  -m_{\isotope[3]{He}} + 2m_\chi-E_e \right)  -m_e^2 \notag \\ 
& \qquad +2 \lambda \left( E_e^2-m_e^2 \right)    +\lambda^2 \left( E_e \left(  2m_{\isotope[3]{H}} +m_{\isotope[3]{He}} +2m_\chi-E_e \right)  -m_e^2 \right) \bigg].
\end{align} 
Unlike in heavy elements where corrections due to the Fermi-Dirac distribution of nucleons in a nucleus are difficult to compute, for tritium, these have been well studied. To account for such corrections, we map $\lambda \to \sqrt{2.788/3} \lambda$~\cite{Schiavilla:1998je}. Then, in the light dark matter limit, we reproduce the standard neutrino capture cross section on tritium~\cite{Long:2014zva}.

\begin{figure}[t!]
\centering \hspace{-0.9cm}
\includegraphics[width =0.78\textwidth]{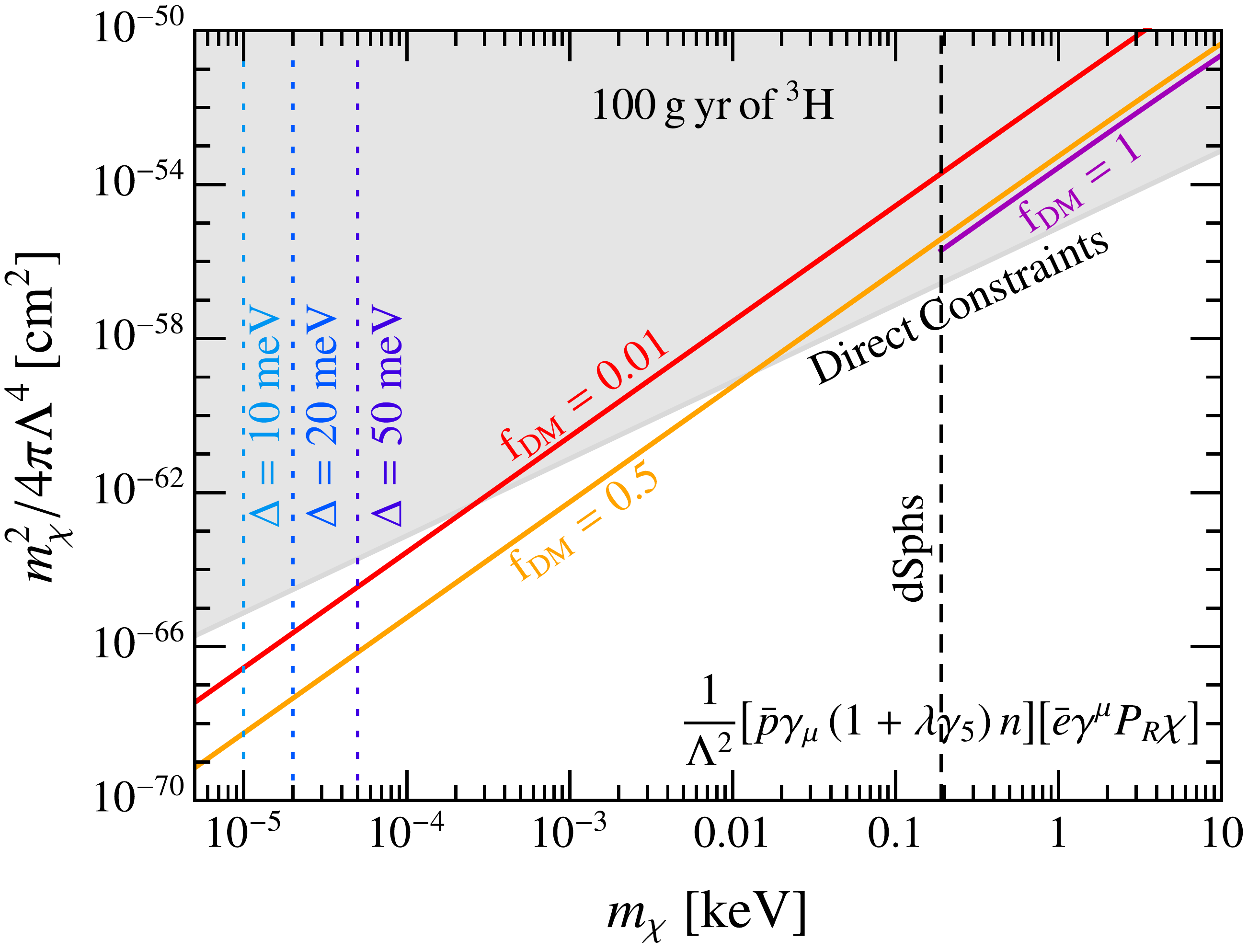}
\caption{\label{fig:H3betaEP} Bounds from dark matter capture events on tritium inside an experiment with a 100 g yr, 1 kg yr, or 10 kg yr exposure. For reference, PTOLEMY has a proposed exposure of at least 100 g yr~\cite{Betti:2019ouf}. Also shown is the LHC bound~\cite{Khachatryan:2014tva} on our UV completion and the lightest possible fermionic dark matter  $m_\chi \sim 190 \text{ eV}$ consistent with dwarf spheroidal galaxies~\cite{DiPaolo:2017geq,Savchenko:2019qnn}.} 
\end{figure}

To project the sensitivity of future tritium-based experiments, we again require the number of absorption events to be less than 10. To connect the projected sensitivities of this low-$m_\chi$ region to those already considered above from the induced $\beta$ signals, we show projected bounds on $\sigma \equiv m_\chi^2 / \prn{4 \pi \Lambda^4}$ in Fig.~\ref{fig:H3betaEP} for tritium exposures of 100 g yr, 1 kg yr, and 10 kg yr. PTOLEMY \cite{Betti:2019ouf} expects to have an exposure of at least 100 g yr. We also show the direct searches bound~\cite{Khachatryan:2014tva} on the UV completion from Sec.~\ref{subsec:CCmodel}. It is interesting that with less than 1~kg yr of tritium, a future experiment could start probing the lightest possible fermionic dark matter~\cite{DiPaolo:2017geq,Savchenko:2019qnn} with these charged current interactions. This provides further motivation to pursue proposals such as PTOLEMY.

\newpage
\section{Discussion}
\label{sec:disc}
In this work, we comprehensively consider signals from the absorption of fermionic dark matter by nuclear targets at direct detection and neutrino experiments. 
These signals arise from a set of dimension-6 operators which do not conserve dark matter number and can be broadly classified into ``neutral current'' and ``charged current'' varieties. We present simple UV completions which lead to these operators and consider bounds from indirect searches for dark matter decays, as well as bounds coming from searches at collider experiments.

The neutral current operators induce dark matter velocity-independent nuclear recoils at distinct energies with relative spacing and peaks which result in a distinguishable signal. We present the general expressions for the rates as well as study the kinematics. We find that future (lower threshold) dark matter experiments employing lighter targets can achieve sensitivity to $m_\chi \lesssim \text{ MeV}$ while remaining consistent with bounds from collider searches and indirect detection. Due to decay rates scaling with large powers of $ m _\chi $, above an MeV, the bounds from indirect detection become stringent. However, these constraints depend on the UV completion, while our projected sensitivities do not, so indirect detection constraints can in principle be fine-tuned away. Regardless, current dark matter and neutrino experiments can similarly probe a large unexplored parameter space. 

In the presence of a dark matter background, the charged current operators can induce $\beta$ decays in otherwise stable isotopes. This yields multiple possible signals: the ejected energetic $e^\pm$, the nuclear recoil of the daughter nucleus, a prompt $\gamma$ from the decay of the excited daughter nucleus, and further decay if its unstable. These correlated signals (unique for every isotope in an experiment) could be searched for simultaneously to reduce possible backgrounds. While one may consider both $ \beta ^- $ and $ \beta ^+ $ decays for any element, large suppressions in $ \beta ^+ $ rates in heavier isotopes makes $ \beta ^- $ the more promising candidate for every element, other than Hydrogen. We project sensitivity for a variety of current dark matter and neutrino experiments, finding powerful sensitivities, easily surpassing current direct and indirect constraints for $ 300~ {\rm keV} \lesssim m _\chi \lesssim 30 ~{\rm MeV}$. In addition, we make projections for induced $\beta^+$ decays in Hydrogen for Borexino and Super-Kamiokande. Due to their shear size, we find these experiments can probe deep into unexplored parameter space, having the largest potential impact for heavier masses.

While induced $ \beta $ decays are prominent signals that can be seen in almost any dark matter or neutrino experiment, they inevitably require dark matter that is sufficiently heavy to induce such a transition putting the rough lower bound on the sensitivities of $ m _\chi \gtrsim 500~ {\rm keV} $. To probe lower masses one can instead look for shifts in the kinematic endpoint of $ \beta$ spectra in isotopes that are already unstable in a vacuum. Due to the lack of existing or future experiments with such targets, we focused on the projected sensitivity of a tritium-based experiment, such as PTOLEMY. We find that such experiments could probe the lightest possible fermionic dark matter consistent with phase space packing bounds which interacts with the SM through these charged current operators.

There is a host of current experiments which could discover dark matter from dedicated analyses for signals from the absorption of fermionic dark matter on nuclear targets. As such, different experiments could probe complementary  unexplored regions of parameter space. The possibility of dark matter which interacts with the SM through either neutral or charged current operators further motivates many proposed future experiments which have other concrete physics goals. As the quest for dark matter leads us away from the WIMP paradigm and into the ocean of light dark matter scenarios, fermionic absorption represents an exciting new class of signals that could, in the near future, discover the nature of dark matter. 

\newpage 
\appendix

\section{Neutral Current Rate at Higher Order}
\label{sec:NCdEdERderiv}

To produce Fig.~\ref{fig:dRdERCRESST}, we need to evaluate the energy-conserving $\delta$ function in Eq.~\eqref{eq:NCdsigma} at $\ord{v^1}$ since evaluating it at $\ord{v^0}$ yields a differential scattering rate proportional to a delta function in $E_R$ (see Eq.~\eqref{eq:NCdRdERatv0}). At $\ord{v^1}$, we find
\al{
\label{eq:NCdeltaatv1}
\delta \prn{E_R+p_\nu-m_\chi \prn{1+\frac{v^2}{2}}} \simeq \frac{\delta \prn{  \cos \theta_{qv} -  \cos \theta_{qv}^0 }}{m_\chi v},
}
where
\al{
\cos \theta_{qv}^0 = \frac{E_R +\sqrt{2 M E_R}-m_\chi}{m_\chi v}.
}
The superscript indicates that this is the value for $\cos \theta_{qv}$ at which the energy-conserving $\delta$ function's argument vanishes. This cosine's allowed range places a minimum condition on $v$:
\al{
\label{eq:NCvmin}
v_{\text{min}}= \frac{\magn{E_R +\sqrt{2 M E_R}-m_\chi}}{m_\chi}.
}

The differential scattering rate at $\ord{v}$ is then
\al{
\frac{dR}{dE_R}= N_T \frac{\rho_\chi}{m_\chi} \sqrt{\frac{E_R}{2M}} \frac{\magn{\MM_N}^2}{16 \pi m_\chi^2 p_\nu} \int d^3v \frac{f(v)}{v} \theta \prn{v-v_{\text{min}}}
}
where $p_\nu=\sqrt{2\sqrt{2}m_\chi \sqrt{E_R M}-2E_R \prn{M+\sqrt{2M E_R}}}$. We approximate the dark matter velocity distribution with a capped Maxwell distribution (see \cite{Lin:2019uvt} for a review)
\al{
f(\vec{v})=\frac{1}{N} \exp \left[- \frac{\prn{\vec{v}+\vec{v}_e}^2}{v_0^2} \right] \theta \prn{v_{\text{esc}}-\magn{\vec{v}+\vec{v}_e}},
}
where $N=\pi^{3/2} v_0^3 \prn{ \text{erf} \left[ v_{\text{esc}}/v_0 \right]-\frac{2 v_{\text{esc}}}{\sqrt{\pi}v_0} \exp \left[- v_{\text{esc}}^2/v_0^2 \right] }$ normalizes the velocity distribution to unity, $v_e \simeq 240 \text{ km/s}$ is the Earth's approximate galactic velocity (dominated by the Sun's), $v_0 \simeq 220 \text{ km/s}$, and $v_\text{esc} \simeq 550 \text{ km/s}$ is the galactic escape velocity.
Thus, the differential scattering rate on a single isotope $j$ per target mass is
\al{
\frac{1}{M_T} \frac{dR_j}{dE_R}= \frac{\rho_\chi}{m_\chi} \sigma_{NC} \frac{N_j M_j \sqrt{2E_R M_j} }{2M_T m_\chi^2 p_\nu} A_j^2F_j^2 \avg{\frac{1}{v}}_{v>v_{\text{min}}},
}
where $v_{\text{min}}$ is given by Eq.~\eqref{eq:NCvmin}. With this, we produce the differential scattering rates from fermionic absorption off the few target isotopes in CRESST in Fig.~\ref{fig:dRdERCRESST}. 

\newpage
\section{Relevant Current Experiments}
\label{sec:appExp}
\begin{table*}
\center
\bgroup
\def\arraystretch{1.2}
\setlength{\tabcolsep}{3pt}
\begin{tabular}{lccccc}
\toprule[1.5pt]
Experiment & Goal & Exposure & Target   & $E_{\text{NR}}^{th}$ &  Refs  \\ 
\midrule[1pt] 
CRESSTII & DM & 52 kg day & $\rm CaWO_4$ crystals & 307 eV & \cite{Angloher:2015ewa} \\
CRESSTIII & DM & 2.39 kg day & $\rm CaWO_4$ crystals & 100 eV & \cite{Petricca:2017zdp} \\
DAMIC & DM & 0.6 kg day & Si CCDs & 0.7 keV &  \cite{Aguilar-Arevalo:2016ndq} \\
DarkSide-50 & DM & 6786 kg day & Liquid Ar & 0.6 keV &  \cite{Agnes:2014bvk,Agnes:2018ves} \\
EDELWEISS & DM & .0334 kg day & Ge & 0.06 keV &  \cite{Armengaud:2019kfj} \\
LUX & DM & 91.8 kg yr & Liquid Xe & 4 keV &  \cite{Akerib:2016vxi} \\
NEWS-G & DM & 9.7 kg day & Neon & 720 eV &   \cite{Arnaud:2017bjh} \\
PandaX-II & DM/$ 0 \nu 2 \beta $ & 150 kg yr & Liquid Xe & 3 keV &  \cite{Cui:2017nnn} \\
PICO-60 & DM & 3420 kg day & Superheated $\rm CF_3I$ & 13.6 keV & \cite{Amole:2015pla} \\
PICO-60 & DM & 1167 kg day & Superheated $\rm C_3F_8$ & 3.3 keV &  \cite{Amole:2017dex} \\
SuperCDMS  & DM & 577 kg day & Ge crystals & 1.6 keV & \cite{Agnese:2014aze} \\
CDMSlite & DM & 70 kg day & Ge crystals & 0.4 keV &  \cite{Agnese:2015nto} \\
XENON1T & DM & 1.0 t yr & Liquid Xe & 3 keV & \cite{Aprile:2018dbl} \\ 
 \midrule[0.5pt]
CUORE & $ 0 \nu 2 \beta $ & 86.3 kg yr & TeO$_2$ crystals & 100 keV &  \cite{Alduino:2017ehq} \\ 
EXO-200 & $ 0 \nu 2 \beta $& 233 kg yr & Liquid $\prescript{136}{54}{\text{Xe}}$  & --- & \cite{Albert:2017qto} \\ 
KamLAND-Zen& $ 0 \nu 2 \beta$  & 504 kg yr &   $\prescript{136}{54}{\text{Xe}}$ in LS & --- &  \cite{Shirai:2017jyz} \\
\midrule[0.5pt] 
Borexino &solar $ \nu $  & 817 t yr &$\rm C_6 H_3 \prn{CH_3}_3$ & 500 keV &  \cite{Agostini:2018fnx} \\ 
COHERENT & CE$\nu$NS & 6726 kg day & CsI[Na] & 6.5 keV &  \cite{Akimov:2017ade,Scholberg:2018vwg}  \\ 
Super-Kamiokande &  $ \nu $& 171,000 t yr & $\rm H_2 O$  & ---  &   \cite{Wan:2019xnl} \\ 
\bottomrule[1.5pt]
\end{tabular}
\egroup
\caption{Experiments which can probe fermionic dark matter absorption signals for which we show projected sensitivities in Figs.~\ref{fig:NCsigma}, \ref{fig:betaminusProjections}, and \ref{fig:betaplusProjections}. Experiments without an explicit nuclear recoil threshold, $E_{\text{NR}}^{th}$, are not used for neutral current projections.}
\end{table*}
Here we summarize the relevant details of all current experiments for which we project sensitivities in Figs.~\ref{fig:NCsigma}, \ref{fig:betaminusProjections}, and \ref{fig:betaplusProjections}. A few additional comments are in order for some of these experiments. We give projections based on Run 2 of CDMSlite \cite{Agnese:2015nto} and not Run 3 \cite{Agnese:2018gze} since Run 2 had a larger exposure and roughly the same threshold. We conservatively underestimate Borexino's exposure by assuming that the 3218 days which had at least an 8-hr exposure only had an 8-hr exposure. Borexino's electron equivalent energy threshold is close to 70 keV \cite{Bellini:2009jr}, which corresponds to a proton recoil threshold of 500 keV \cite{Tretyak:2013xta}. It's also worth noting that the Carbon recoil threshold is too high thanks to its poor relative light yield. EXO-200 is $80 \%$ $\prescript{136}{54}{\text{Xe}}$, while KamLAND-Zen is $91 \%$.

\acknowledgments
We thank Artur Ankowski, Carlos Blanco, Tim Cohen, Jack Collins, Simon Knapen, Tongyan Lin, Ian Moult,  Maxim Pospelov, and Lorenzo Ubaldi for useful discussions, and Jason Detwiler, Volodymyr Tretyak, Vetri Velan, and Lindley Winslow for input on the capabilities of dark matter detectors. JD is supported in part by the DOE under contract DE-AC02-05CH11231.  GE is supported by the U.S. Department of Energy Award DE-SC0011637. 
\bibliographystyle{JHEP}
\bibliography{Fabs}

\end{document}